\documentclass[aps, prb, twocolumn, showpacs, groupedaddress, showkeys]{revtex4}
\usepackage[dvips]{color}
\usepackage{amsfonts}
\usepackage{amsmath}
\usepackage{amssymb}
\usepackage{graphicx}
\usepackage{mathrsfs}
\usepackage{bbm}
\usepackage{bm}
\usepackage{multirow}
\usepackage[sort&compress]{natbib}
\usepackage{dcolumn}%

\begin{document}

\newcommand{\dd}{d}
\newcommand{\pd}{\partial}
\newcommand{\myU}{\mathcal{U}}
\newcommand{\myr}{q}
\newcommand{\Urho}{U_{\rho}}
\newcommand{\myalpha}{\alpha_*}
\newcommand{\bd}[1]{\mathbf{#1}}
\newcommand{\Eq}[1]{Eq.~(\ref{#1})}
\newcommand{\Eqn}[1]{Eq.~(\ref{#1})}
\newcommand{\Eqns}[1]{Eqns.~(\ref{#1})}
\newcommand{\Figref}[1]{Fig.~\ref{#1}}
\newtheorem{theorem}{Theorem}
\newcommand{\me}{\textrm{m}_{\textrm{e}}}
\newcommand{\sgn}{\textrm{sign}}
\newcommand*{\bfrac}[2]{\genfrac{\lbrace}{\rbrace}{0pt}{}{#1}{#2}}

\newcommand{\CTensorName}{transport shape-averaged }

\title{Theory of Band Warping and its Effects on Thermoelectronic Transport Properties}

\author{Nicholas A. Mecholsky}
 \email{nmech@vsl.cua.edu}
 \homepage{http://ape.umd.edu}
 \thanks{Corresponding Author}

\author{Lorenzo Resca}
 \email{resca@cua.edu}
 \homepage{http://physics.cua.edu/people/faculty/resca.cfm}

\author{Ian L. Pegg}
 \email{ianp@vsl.cua.edu}

\affiliation{
Department of Physics and Vitreous State Laboratory\\
The Catholic University of America\\
Washington, DC 20064}

\author{Marco Fornari}
 \email{marco.fornari@cmich.edu}
 \homepage{http://www.phy.cmich.edu/people/fornari/}

\affiliation{
Department of Physics\\
Central Michigan University\\
Mount Pleasant, Michigan 48858}

\date{\today}

\begin{abstract}
Optical and transport properties of materials depend heavily upon features of electronic band structures in proximity of energy extrema in the Brillouin zone (BZ). Such features are generally  described in terms of multi-dimensional quadratic expansions and corresponding definitions of effective masses. Multi-dimensional quadratic expansions, however, are permissible only under strict conditions that are typically violated when energy bands become degenerate at extrema in the BZ. Even for energy bands that are non-degenerate at critical points in the BZ there are instances in which multi-dimensional quadratic expansions cannot be correctly performed. Suggestive terms such as ``band warping'', ``fluted energy surfaces'', or ``corrugated energy surfaces'' have been used to refer to such situations and \textit{ad hoc} methods have been developed to treat them. While numerical calculations may reflect such features, a complete theory of band warping has not hitherto been developed. We define band warping as referring to band structures that do not admit second-order differentiability at critical points in $\bd{k}$-space and we develop a generally applicable theory, based on radial expansions, and a corresponding definition of angular effective mass. Our theory also accounts for effects of band non-parabolicity and anisotropy, which hitherto have not been precisely distinguished from, if not utterly confused with, band warping. Based on our theory, we develop precise procedures to evaluate band warping quantitatively. As a benchmark demonstration, we analyze the warping features of valence bands in silicon using first-principles calculations and we compare those with previous semi-empirical models. As an application of major significance to thermoelectricity, we use our theory and angular effective masses to generalize derivations of tensorial transport coefficients for cases of either single or multiple electronic bands, with either quadratically expansible or warped energy surfaces. From that theory we discover the formal existence at critical points of transport-equivalent ellipsoidal bands that yield identical results from the standpoint of any transport property. Additionally, we illustrate with some basic multi-band models the drastic effects that band warping and anisotropy can induce on thermoelectric properties such as electronic conductivity and thermopower tensors.
\end{abstract}

\pacs{72.10.Bg, 02.30.Mv, 72.20.Pa, 71.18.+y, 72.80.Cw}
\keywords{band warping, thermoelectricity, Boltzmann equation, electronic transport}

\maketitle

\section{Introduction}

Basic aspects of optical and transport properties in materials such as metals, semiconductors, and insulators can be described in terms of elementary approximations based on the so called free (or nearly free) electron model.\cite{drude1900electronI, drude1900electronII, lorentz1916theory, sommerfeld1928elektronentheorie, bloch1929quantenmechanik, Ashcroft, Kittel, Marder, Bassani, GP, Singleton, Bastard, SchaferWegener2002}  Although predictions derived from free electron models may not be quantitatively accurate, they provide critical insight into physical processes underlying many situations, such as optoelectronic phenomena and diffusive electronic transport in nano-structures, mesoscopic systems, and so on.\cite{singh2001theoretical, jeong2012best, wang2011assessing} Searching for optimal thermoelectric materials, the guiding principles that refer to band structures are usually formulated in terms of parameters of the free electron model, which enter expressions of the electronic conductivity, $\underline{\bm{\sigma}}$, thermopower, $\underline{\bm{S}}$, and other transport-coefficient tensors and figures of merit.\cite{Mahan1996, Mahan1989, Curtarolo2010, Snyder2008complex, NolasSharpGoldsmid}

A central concept of the free electron model is that of effective mass. Physically, effective masses are parameters that originate from strong interactions, but end up producing ``quasi-particles'' that move freely for some energies and directions.\cite{landau1956sov, landau1957oscillations, Kaxiras} Mathematically, effective masses may arise from quadratic expansions of dispersion relations, such as those of energy and quasi-momentum in electronic band structures. Those expansions are typically performed around band extrema, although the absence of linear terms is not essentially required.\footnote{Alternative definitions of effective masses including linear terms in the quasi-momentum expansion of the energy band may be considered.\cite{Ariel} That may yield a proportionality between effective mass and quasi-momentum, as observed in graphene, for example. Notwithstanding the interest in that, we shall confine our considerations to the exclusion of linear terms for the treatment of effective masses in most of this paper.} One should carefully reconsider, however, the permissibility of performing multi-dimensional Taylor polynomial expansions of electronic band energies around local extrema, occurring at one or more $\bd{k}_0$ vector points in quasi-momentum space or the Brillouin Zone (BZ). Physicists are accustomed to presume that multi-dimensional Taylor expansions should generally be permissible in physical realizations, perhaps because they seldom encounter instances of mixed partial derivatives that do \textit{not} commute. However, that is precisely a characteristic of ``band warping'' in some frames. In mathematics, that characteristic in fact applies to most ordinary functions of several variables. Some examples are provided in Appendix\ \ref{appendix:Taylor}, in the context of a brief summary of the basic statements and mathematical theorems on Taylor expansions that are needed for a precise understanding of band warping. By that we mean that band warping is the manifestation of a band structure that is not second-order differentiable at a critical point in $\bd{k}$-space.

Most literature in condensed matter physics presumes the existence of multi-dimensional quadratic expansions of energy bands around critical points: see, for instance, textbook examples such as those of Eqs.~(12.29), p.\ 228, and Eqs.~(28.3), p.\ 568, of Ref.~\onlinecite{Ashcroft}; Eqs.~(53), p.\ 71, of Ref.~\onlinecite{GP}; Eq.~(5.25), p.\ 156, of Ref.~\onlinecite{Bassani}; Eq.~(21.10), p.\ 636 of Ref.~\onlinecite{Marder}; Eq.~(3.79), p.\ 69 of Ref.~\onlinecite{SchaferWegener2002}. ``Band warping'' or equivalent effects are incidentally mentioned or separately treated as special cases. Originally, those features were theoretically foreseen and experimentally investigated in seminal papers.\cite{LuttingerKohn1955Motion, Dresselhaus, Kane1957band} Since then, other authors have dealt with band warping in various ways.\cite{Bastard, Lawaetz1971, Boykin1999, Helmholz2002}.

It turns out that energy bands that become degenerate at extrema in the BZ are almost invariably warped. Even energy bands that are non-degenerate at critical points in the BZ have recently been found to be ``corrugated''.\cite{ParkerPRL2013, Chen2013} No multi-dimensional quadratic expansion can be correctly applied to any of those cases, as we shall demonstrate, and an alternative approach is required.

Let us start by recalling the conventional three-dimensional quadratic expansion of an energy band $E_n(\bd{k})$ around a local extremum $\bd{k}_0$ in the BZ, 
\begin{equation}\label{eqn:quad}
E(\bd{k}) \simeq E_0 +  \frac{\hbar^2}{2 \me} \bd{k} \cdot \underline{\bm{M}}^{-1} \cdot  \bd{k},
\end{equation}
where $\me$ represents the electron mass and $E_0 = E(\bd{k}_0)$. The inverse effective mass tensor $\underline{\bm{M}}^{-1}$ is defined by the Hessian matrix of second-order partial derivatives computed at $\bd{k}_0$ as
\begin{equation}\label{eqn:mass}
\left[ \underline{\bm{M}}^{-1} \right]_{ij} = \left. \frac{\me}{\hbar^2} \frac{\pd^2 E_n(\bd{k})}{\pd k_i \pd k_j} \right|_{\bd{k}_0}.
\end{equation}
However, \Eqns{eqn:quad} and (\ref{eqn:mass}) are inapplicable to multi-variable functions that are not second-order differentiable in a mathematically rigorous sense. Those represent cases of ``band warping'', according to the terminology of most physicists, which we may retain, as long as our rigorous definition is understood. 

The more general theory that we develop in this paper consists in performing a one-dimensional Taylor polynomial expansion in any fixed radial direction, originating at the band extremum. Mathematically, that requires only one-dimensional differentiability to any given order $n$ at the expansion point. We thus expect that most physical band energy structures should admit such one-dimensional expansions in any preset radial direction originating at a band extremum. This basic consideration lies at the heart of our method of formulating an angular-effective-mass approximation that is rigorous and much more general than that of the conventional multi-dimensional-expansion approach. The set of quadratic-term coefficients in all the radial expansions defines the angularly-dependent effective mass. This approach thus allows an automatically correct treatment of band warping. It also avoids incorrect multi-dimensional ``integration by parts'' (Green's Theorem) based on \Eqns{eqn:quad} and (\ref{eqn:mass}): see, for instance, Eq.~(13.27) on p.\ 251 of Ref.~\onlinecite{Ashcroft}, incorrectly developing the electric conductivity tensor from the more general and correct definition given in Eq.~(13.25) on p.\ 250.

We shall also consider the set of cubic-term coefficients and those of even higher order \textit{n} in the radial expansions for the energy band dispersion around a local extremum. Those angularly-dependent coefficients become increasingly responsible for \textit{radial band non-parabolicity} effects as we move further away from the extremum. The one-dimensional nature of the radial expansions causes no particular problem for such \textit{radial} treatment of band non-parabolicity. In contrast, far more restrictive conditions would have to apply to \textit{multi-dimensional} Taylor polynomial expansions of higher order.

The outline of our paper is as follows. In Sec.\ \ref{angmass} we provide a mathematically precise definition of warping in electronic band-structure theory, and in Sec.\ \ref{sec:warping} we develop a general theoretical and computational procedure to quantify warping in band structures. In Sec.\ \ref{sec:silicon} we demonstrate those procedures with an application to silicon, using both first-principles calculations and $\bd{k} \cdot \bd{p}$ approximations to degenerate and non-degenerate bands at a critical point, discussing in particular the origin and interplay of band non-parabolicity and warping. In Sec.\ \ref{trancoef} we formulate a microscopic Boltzmann-Onsager theory of transport in anisotropic materials to include both warped and non-warped multi-band and multi-valley structures, thus generalizing previous treatments, such as that of Ref.~\onlinecite{Bies2002}, which retain the unwarranted assumption of multi-dimensional quadratic expansions and single bands. In Sec.\ \ref{sec:equivalentellipsoid} we report the discovery of the formal existence of equivalent-transport ellipsoids for warped bands. In Sec.\ \ref{undewar} we demonstrate drastic effects that band warping can produce on electronic conductivity and thermopower tensors under certain conditions. Physical effects of that kind have been recently observed by various authors.\cite{ParkerPRL2013, Chen2013, Usui2010, Filippetti2012, Shirai2013} In Sec.\ \ref{Conclusion} we summarize our conclusions regarding how band warping, band non-parabolicity, anisotropy, multi-valley and multi-band structures can affect and possibly optimize electronic transport for thermoelectricity.  

\section{The angular effective mass}\label{angmass}

Our theory is based on the general assumption that, in an open neighborhood of $\bd{k}_0$ within the BZ, the energy-momentum dispersion relation for a given band can be expressed in polar coordinates and then expanded for fixed $\theta$ and $\phi$ as  
\begin{align}\label{eqn:generalTaylor}
E(k_r,\theta, \phi) =& E_0 + a_1(\theta, \phi) k_r \nonumber\\&+ a_2(\theta, \phi) k_r^2 + a_3(\theta, \phi) k_r^3+ \ldots
\end{align}
The basic idea is to perform a one-dimensional Taylor series expansion in any preset radial direction originating at $\bd{k}_0$, with coefficients that depend parametrically on the polar angles identifying that given radial direction. It is reasonable to assume that physical band energy-momentum dispersion relations typically satisfy one-dimensional analyticity requirements in every specified angular direction. Then each radial Taylor series expansion must converge to $E(k_r,\theta, \phi)$ in an open interval surrounding $\bd{k}_0$ in each fixed angular direction and its opposite. Let us stress again the fact that \Eq{eqn:generalTaylor} represents a far broader class of functions than those limited by multi-dimensional Taylor expansions like \Eq{eqn:quad} and beyond. \Eq{eqn:generalTaylor} thus allows for the possibility of including band warping, radial non-parabolicity, and other features in our physical descriptions.

For the moment, let us limit our radial expansions to second-order, while higher-order terms resulting in radial band non-parabolicity will be considered later. Presently, linear terms may also be excluded, considering special points in the BZ where they must vanish as a result of time-reversal or other symmetry requirements. We thus obtain from \Eq{eqn:generalTaylor} that
\begin{equation}\label{eqn:deff}
E(k_r,\theta, \phi) \simeq E_0 + \frac{\hbar^2 k_r^2}{2 \me} f(\theta, \phi),
\end{equation} 
where the coefficient $\frac{\hbar^2}{2 \me}$ is again introduced in conformity with standard conventions. Thereby,
\begin{equation}\label{eqn:effmass}
f(\theta,\phi) = \frac{1}{m(\theta,\phi)}
\end{equation}
represents a dimensionless ``angular effective mass surface'', where explicit reference to inversion may be omitted for brevity. Setting the coefficient $\frac{\hbar^2}{2 \me}=1$ implies selection of Rydberg atomic units, where $\hbar=2 \me=\frac{e}{\sqrt{2}}=1$. For convenience of units, the same coefficient $\frac{\hbar^2}{2 \me}$ will also be introduced later in \Eq{eqn:kittel}, in comparison with its original form.\cite{Dresselhaus}
 
Equations (\ref{eqn:deff}) and (\ref{eqn:effmass}) are both mathematically and physically far more general than \Eqns{eqn:quad} and (\ref{eqn:mass}). Contrary to the latter equations, they inherently account for the possibility of band warping. That possibility is instead excluded in \Eqns{eqn:quad} and (\ref{eqn:mass}) by their much more restrictive assumption of multi-dimensional second-order differentiability.\footnote{There are, of course, classes of mathematical functions that may not admit one-dimensional Taylor series expansions in all radial directions originating at $\bd{k}_0$, contrary to the assumption of \Eq{eqn:generalTaylor}. However, as we discussed in that regard, those functions are unlikely to represent physically meaningful electronic band structures. Thus, in practice, we may restrict the class of all physically meaningful band-warped functions to those that can be expanded as in \Eq{eqn:deff} at a critical point.}

There may be additional symmetry constraints on the angular dependence of $m(\theta,\phi)$, but those are completely justified and of an entirely different nature, originating from the possibility of a finite number of discrete symmetry operations belonging to the small point group of $\bd{k}_0$: see, for instance, Refs.~\onlinecite{Bassani, Parmenter55, DresselhausTerm, Phillips56}. 

Assumptions similar to ours have been previously made, stemming from considerations that effective masses should be physically defined and measurable for every possible direction around critical points in the BZ: see, for instance, Refs.~\onlinecite{Dresselhaus}, \onlinecite{Ottaviani1975}, and \onlinecite{JacoboniReggiani1983}. A seminal paper by Phillips is particularly valuable from our perspective, since it deals with lattice vibration spectra at critical points that also display warping effects.\cite{Phillips56} However, \textit{a full angularly-dependent theory of lattice vibrations needs much further consideration}.

It is easier to appreciate the superiority of the angular effective mass approach by considering at first some elementary applications. One that we generalize later to \Eq{eqn:kittel} represents the prototypical situation of top valence bands in cubic semiconductors investigated originally by Dresselhaus, Kip, and Kittel.\cite{Dresselhaus} Namely, consider a two-dimensional band-structure having
\begin{equation}\label{eqn:toymodel}
E(k_x, k_y) = \frac{\hbar^2}{2 \me}\sqrt{k_x^4 + k_y^4}.
\end{equation}
This function is 1st-order differentiable everywhere and it has a minimum at the origin. However, its Hessian matrix of second-order partial derivatives is discontinuous at the zone center. Hence, $E(k_x, k_y)$ is not second-order differentiable at (0,0), admitting no valid two-dimensional quadratic expansion therein. In other words, as noted in Appendix\ \ref{appendix:Taylor}, we may formally compute an apparently adequate Hessian matrix at the origin, but that cannot possibly transform as it should under rotations of the coordinate axes. Notwithstanding that, when expressed in polar coordinates ($k_x = k_r \cos \theta$, and $k_y = k_r \sin \theta$) as
\begin{equation}\label{eqn:toymodelangular}
E(k_r, \theta) = \frac{\hbar^2}{2 \me} k_r^2 \frac{1}{2} \sqrt{3 + \cos (4\theta)},
\end{equation}
it is apparent that this function is perfectly parabolic in $k_r$ along each radial direction with a given polar angle $\theta$, which defines a corresponding angular effective mass through $f(\theta)$. This provides an elementary physical example of a function which is not analytic at the origin in two dimensions, but it is analytic in one dimension along each given radial direction. Even though this function has no valid Taylor polynomial approximation beyond first order in two dimensions, let alone any valid two-dimensional Taylor series, its angular effective mass $f(\theta)$ is perfectly well defined and in fact it provides an exact description of the function in the entire space.

\section{Quantitative Measures of Band Warping}\label{sec:warping}

Previous authors have attempted to provide measures of band warping, using certain methods of band structure calculations and \textit{ad hoc} definitions for certain materials or considering limited sections of the BZ.\cite{Lawaetz1971, Helmholz2002} Our angular effective mass approach enables a more general formulation in all those respects.
 
Since the construction of Hessian matrices depends critically on the choice of coordinate axes, we may exploit any variance of eigenvalues of Hessian matrices formally computed in different coordinate systems to reveal the presence of band warping and generate corresponding measures of its magnitude.  When there is no band warping, hence we have a valid multi-dimensional quadratic expansion and a corresponding ``bona fide'' Hessian matrix, its eigenvalues must be real and invariant under orthogonal transformations or, equivalently, independent of any choice of Cartesian coordinate axes. Now, even when there is band warping, we may typically be able to compute second-order partial derivatives at an extremum in any given coordinate system, because second-order partial derivatives require only one-dimensional limits, however performed in a double sequence. From those partial derivatives we may thus formally construct a (mala fide) Hessian matrix, and we may do likewise in any other Cartesian coordinate system. However, the corresponding Hessian matrices will not properly transform into one another according to the appropriate orthogonal coordinate transformations, nor will their eigenvalues necessarily remain real and invariant as they should: cf.\ the examples mentioned in Appendix\ \ref{appendix:Taylor}. We may exploit that failure by constructing corresponding measures of band warping as follows.

First take the angular average of the traces of Hessian matrices, corresponding to the sum of their eigenvalues in each coordinate system. Then take the angular average of the square of the difference between the Hessian trace in any given coordinate system and the prior averaged trace. The square root of the latter angular integration gives a root-mean-square (RMS) value. That RMS deviation can be further divided by the original angularly averaged trace, thus providing a coefficient of variation (COV) of Hessian matrix traces averaged over all radial directions. We may regard those RMS and COV as absolute and relative measures of band warping, respectively. 

A dimensionless warping parameter $w$, representing that COV, may thus be defined as
\begin{equation}\label{eqn:COV}
w = \frac{\langle  \, \, \left( Tr[H_\Omega] - \langle Tr[H_\Omega] \rangle \right)^2 \, \, \rangle^{1/2}}{\langle Tr[H_\Omega] \rangle},
\end{equation}
where $\textrm{Tr}[H_\Omega] = \sum_{i=1}^{D} e_i(\Omega)$ is the trace and $e_i$ are the eigenvalues of the Hessian matrix\footnote{Since the Hessian matrix is a real matrix, its trace must be real, although its eigenvalues may be complex.} and $D$ equals 2 or 3 for two or three dimensions. Orientation of the axes is defined by a set of Euler angles $\Omega = \{ \Phi,\Theta,\Psi \}$ in three dimensions, or by a single angle $\Omega = \{ \theta \}$ in two dimensions. The angular average is defined by
\begin{equation}
\langle \cdot \rangle_{\theta} = \frac{1}{2 \pi} \int_{0}^{2 \pi} \cdot \, \, \,  \dd \theta
\end{equation}
in two dimensions, and by
\begin{equation}\label{eqn:COVin3D}
\langle \cdot \rangle_{\Phi,\Theta,\Psi} = \frac{1}{8 \pi^2} \int_{0}^{2 \pi}  \! \!  \int_{0}^{\pi} \! \! \int_{0}^{2 \pi} \cdot \, \, \sin (\Theta) \,  \, \dd \Phi \,  \dd \Theta  \, \dd \Psi
\end{equation}
in three dimensions.

A close estimate of \Eq{eqn:COV} may be obtained from a randomly chosen collection of coordinate directions by performing the angular average
\begin{equation}
\langle \cdot \rangle_{n} = \frac{1}{n} \sum_{i=1}^{n} \, \cdot 
\end{equation}
in terms of $n$ randomly sampled coordinate axes orientations.

When we have an energy dispersion set in the form of \Eq{eqn:deff}, we can express the trace of Hessian matrices in dimensionless form simply as a function of the angular effective mass, namely
\begin{subequations}\label{eqn:Traces}
\begin{align}
\textrm{Tr}[H(\theta)] &= 2 f(\theta) + 2 f(\theta + \pi/2), \\
\textrm{Tr}[H(\Phi,\Theta,\Psi)] &= \sum_{i=1}^3 2 f(\theta_i(\Phi,\Theta,\Psi), \phi_i(\Phi,\Theta,\Psi)),
\end{align}
\end{subequations}
in 2D and 3D, respectively. In 3D, the polar angles $ ( \, \theta_i(\Phi,\Theta,\Psi), \phi_i(\Phi,\Theta,\Psi) \, )$ of the $i^{\textrm{th}}$ axis are given by

\begin{align}
\theta_i (\Phi,\Theta,\Psi) &= \tan^{-1}( \mathbb{A}_{3,i} , \sqrt{\mathbb{A}_{1,i}^2 + \mathbb{A}_{2,i}^2 } ) ,\\
\phi_i (\Phi,\Theta,\Psi) &= \tan^{-1}( \mathbb{A}_{1,i} , \mathbb{A}_{2,i} ) ,
\end{align}
where $\mathbb{A}$ is the matrix whose columns define the coordinate axes relative to a fixed reference system, and $\tan^{-1}(x,y)$ gives an angle between $0$ and $2 \pi$ that respects the signs of $x$ and $y$. Here $i$ can be 1, 2, or 3, referring to the $x$, $y$, and $z$ axes, respectively. 

It is relatively easy to see why in \Eq{eqn:Traces} all traces of Hessian matrices can be expressed exclusively in terms of angular effective mass surfaces, without involvement of their derivatives. One may recall that, by construction, the angular effective mass function represents half of the double directional derivative along each radial direction. Correspondingly, the trace of a Hessian matrix represents the sum of the second derivatives along each coordinate axis.

Given these definitions, the simple example of $g(x,y) = \sqrt{x^4 + y^4}$, mentioned in the Appendix\ \ref{appendix:Taylor} and in \Eq{eqn:toymodel}, has a warping parameter $w = \tfrac{\sqrt{3 \pi^2 - 16 \textrm{K}(1/2)^2}}{4\textrm{K}(1/2)} \approx 0.1201$, where $\textrm{K}$ denotes the complete elliptic integral of the first kind. Separately, the RMS or standard deviation in the numerator of \Eq{eqn:COV} has a value of $2 \sqrt{ 3 \pi^2 - 16 \textrm{K}(1/2)^2 } / \pi\approx 0.413$ for this simple two-dimensional example. For its straightforward generalization to three dimensions, i.e., $g(x,y,z) = \sqrt{x^4 + y^4 + z^4}$, the warping parameter decreases a little to $w \approx 0.1101$, while the numerator of \Eq{eqn:COV} increases more conspicuously to about 0.5064.

The definition of $w$ stems from the fact that the trace of a bona fide Hessian matrix for a valid multi-dimensional quadratic expansion must be invariant under orthogonal transformations, representing the sum of the invariant real eigenvalues for its symmetric quadratic form. The determinant for that quadratic form, representing the product of its eigenvalues, must also be an invariant, and that may provide a basis for an alternative definition of band warping, stemming from incorrect transformations of corresponding mala fide Hessian matrices.\footnote{In the framework and language of differential geometry for more general curvilinear coordinate transformations, those two alternative approaches to define warping would correspond to quantifying deviations from the expected invariance of mean curvature and intrinsic curvature in smooth manifolds, respectively.} 

\section{Application to Silicon}\label{sec:silicon}
We may now illustrate the significance of the angular effective mass approach with a physical application to realistic bands in silicon, some degenerate at the BZ center ($\Gamma$) and some not. This will demonstrate an improvement in accuracy gained by fitting an actual band structure, calculated from first principles using the $\textsc{Quantum Espresso}$ ($\textsc{QE}$) package.\cite{QE-2009} We performed density functional theory (DFT) calculations where basis functions and BZ integrations were well converged. We used both relativistic and scalar-relativistic norm-conserving Perdew-Burke-Ernzerhof (PBE) pseudopotentials.\cite{QE-2009} We then obtained the angular effective mass surfaces for various bands radially expanded around $\Gamma$ in the BZ.

Considering the top valence bands at $\Gamma$, we fitted \Eq{eqn:deff} to first-principles calculations performed in 2653 different angular directions, distributed uniformly on a hemisphere. Fifteen different points between end-points of $k_r = \pm 0.05 \, \frac{2 \pi}{a_0}$ were evaluated for each energy band and each angular direction, and those fifteen points were fitted to a parabola. We have then systematically assessed the goodness of those parabolic fits. For example, the goodness of radial parabolic fits to the so-called heavy-hole band is reflected by average $R^2$ values of $0.9954 \pm 0.0001$ or $0.999873 \pm 2 \times 10^{-6}$ for computations including spin-orbit interactions or disregarding spin-orbit interactions, respectively (corresponding error estimates are standard errors in the mean). For each particular $(\theta,\phi)$ angular direction, the shape of the fitting parabola generates the function that determines the angular effective mass, namely $f(\theta,\phi)$. For verification, we have systematically reduced the fitting region from $0.05 \, \frac{2\pi}{a_0}$ to $0.001 \, \frac{2 \pi}{a_0}$, which allowed us to evaluate quantitatively the deviation associated with band non-parabolicity. We found none to be significant for the heavy-hole band, but the situation for other bands is more complex.

\begin{figure}[ht!]
	\begin{center}
 	\includegraphics[width=8.6cm]{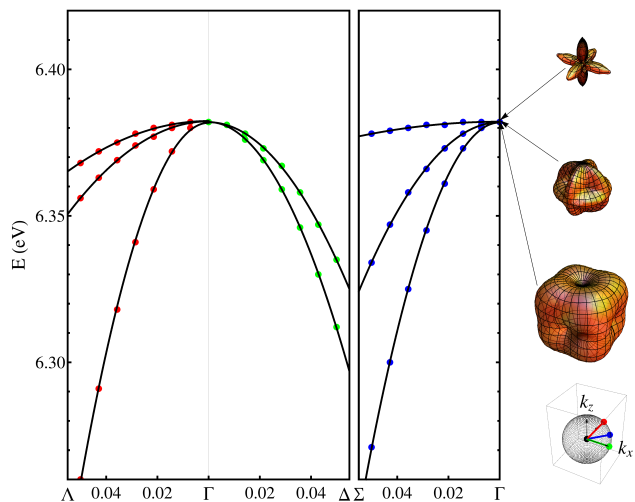}\\
      \caption{\label{angmassSiNSO}(Color) Selected band plots for the top three valence bands of Si near the $\Gamma$ point disregarding spin-orbit interactions.  Red points represent the energy dispersion calculated according to first principles (QE) along the $\Lambda$ line starting at $0.05 \, \frac{2 \pi}{\sqrt{3} a_0}(1,1,1)$ and moving toward $\Gamma$.  Green points move from $\Gamma$ out to $0.05 \, \frac{2 \pi}{a_0}(1,0,0)$ along the $\Delta$ line. Blue points are plotted from $0.05 \frac{2 \pi}{\sqrt{2} a_0}(1,1,0)$ along the $\Sigma$ line toward $\Gamma$. The inset to the bottom right shows the three paths in $\bd{k}$-space selected for the left panels. The solid curves in the left panels represent the radial parabolic approximations to the energy dispersions for each band. Every direction requires a different parabolic fit for each band, as a result of its warping. The curvatures of all the radial parabolas then generate the corresponding non-spherical angular effective mass surfaces shown in the right panel. The top two surfaces correspond to the heavy-hole and light-hole bands.  Without spin-orbit coupling, those two bands are degenerate with a third valence band at $\Gamma$. As a result, all three effective mass surfaces are warped and non-spherical, particularly for the heavy-hole band.}
	\end{center}
\end{figure}

\begin{figure}[ht!]
	\begin{center}
 	\includegraphics[width=8.6cm]{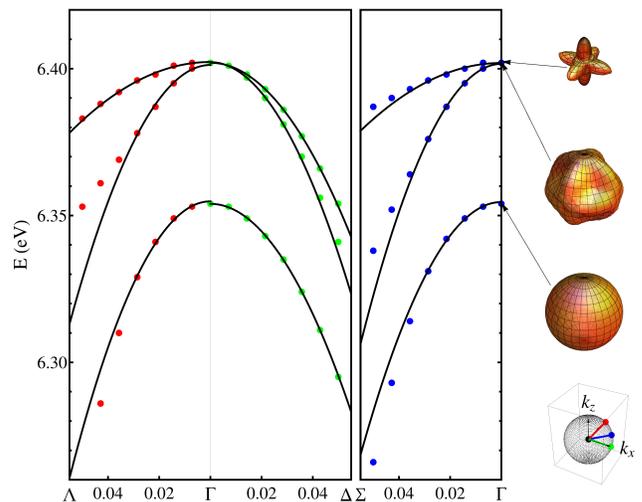}\\
      \caption{\label{angmassSiSO} (Color) Selected band plots for the top three valence bands of Si near the $\Gamma$ point including spin-orbit interactions. Curve and surface descriptions and definitions are the same for all panels as those given in the captions of \Figref{angmassSiNSO}. The top two angular effective mass surfaces shown in the right panel correspond to the heavy-hole and light-hole bands, which remain degenerate at $\Gamma$. As a result, the effective mass surfaces for the top two bands remain warped and non-spherical, particularly for the heavy-hole band. However, the split-off band becomes non-degenerate on account of spin-orbit interactions. It thus appears to lose its warping, as shown by its spherical effective mass surface in the third subgraph down in the right panel.}
	\end{center}
\end{figure}

We show energy band structures and their corresponding angular effective mass surfaces for the top three valence (p-like bonding) bands of Si at $\Gamma$, first disregarding spin-orbit interactions in \Figref{angmassSiNSO}, and then including spin-orbit interactions in \Figref{angmassSiSO}. These energy bands are graphed along the symmetry lines $\Lambda$, $\Delta$, $\Sigma$ originating from $\Gamma$ in the the left two panels of each figure. The actual energy values calculated from QE are plotted as red points along the $\Lambda$ line, green points along the $\Delta$ line, and blue points along the $\Sigma$ line.  The corresponding best parabolic fits are shown as black curves.  For a general band, the $k_r$ interval for a good quadratic approximation in any given angular direction may depend on that direction.  For the heavy-hole band, we have already mentioned that the $0.05 \frac{2 \pi}{a_0}$ range is quite adequate for all angular directions. However, for other bands, the $k_r$ range that allows an adequate quadratic fit may be much smaller or vary more abruptly with angular direction, as can be seen in \Figref{angmassSiSO} already.  In both \Figref{angmassSiNSO} and \Figref{angmassSiSO}, end-points in $\bd{k}$-space for $\Lambda$, $\Delta$, $\Sigma$ have been chosen at $0.05 \, \frac{2 \pi}{\sqrt{3} a_0}(1,1,1)$, $0.05\, \frac{2\pi}{a_0}(1,0,0)$, $0.05 \frac{2 \pi}{\sqrt{2} a_0}(1,1,0)$, for $\Lambda$, $\Delta$, $\Sigma$ lines, respectively. Those end-points are also shown in the inset depicting the corresponding sphere in $\bd{k}$-space at the bottom right of each figure.

The angular effective mass surfaces $f(\theta,\phi)$ associated with each band are shown in the the right panels of \Figref{angmassSiNSO} and \Figref{angmassSiSO}. Notice the qualitative differences in $f(\theta,\phi)$ surfaces generated by splitting of the lowest band by spin-orbit coupling. That splitting eliminates band warping and non-sphericity for the split-off band as a result of removing its degeneracy at $\Gamma$ with the other two higher bands, which instead remain degenerate and warped.

We should recall once more that $f(\theta, \phi)$ surfaces are independent of $k_r$ by definition. Hence, their shape and warping should remain unchanged no matter how close one may attempt to approach $\Gamma$, e.g., by reducing $k_r$ end-points along any angular direction. This contrasts with the hitherto common usage of constant energy surfaces, which vary on a wide scale, depending on the parametrically chosen values of the energy. That makes it much harder to infer the presence of band warping, let alone determine any quantitative estimate for it.

Simple analytical models indicate a typical behavior of two or more interacting non-degenerate non-warped bands when a gap between them is parametrically narrowed and eventually closed.\cite{Resca} The coefficients of the multi-dimensional Taylor series expansions for the non-degenerate bands begin to diverge as the gap between the bands closes and discontinuously transform into band warping exclusively at the point where the bands make contact with each other. First-principles calculations confirm that type of behavior. However, even energy bands that do \textit{not} become degenerate at critical points can occasionally warp or ``corrugate'' through alternative mechanisms, as has been recently demonstrated for the conduction band minima of GeTe and SnTe at L-points.\cite{ParkerPRL2013, Chen2013} \textit{Therefore, the validity of multi-dimensional Taylor expansions at any point must be questioned for energy bands of any type}.

The three-dimensional quadratic expansion was originally questioned for the top valence bands of silicon and germanium in seminal papers.\cite{LuttingerKohn1955Motion, Dresselhaus, Shockley} Energy bands were calculated with a perturbative $\bd{k} \cdot \bd{p}$ method, arriving at the expression\cite{Dresselhaus}
%
\begin{align}\label{eqn:kittel}
&E(\bd{k}) = \nonumber\\ &\frac{\hbar^2}{2 \me} \left( A k^2 \pm \left[ B^2 k^4 + C^2 (k_x^2 k_y^2+ k_y^2 k_z^2 + k_z^2 k_x^2) \right]^{1/2} \right)
\end{align}
for the top two valence bands degenerate at $\Gamma$, after inclusion of spin-orbit interactions and the corresponding degeneracy removal of a split-off lower band.\footnote{The expression corresponding to our \Eq{eqn:kittel} without its pre-factor is first reported in the Abstract and then derived as Eq.\ (63) in Ref.~\onlinecite{Dresselhaus}. A subsequent repetition of that expression in Eq.\ (78) of Ref.~\onlinecite{Dresselhaus} contains a typographical error, since it omits squaring the last $k_z$ in the expression.} 

Using polar coordinates and our definition of angular effective mass function, \Eq{eqn:deff}, we can recast \Eq{eqn:kittel} equivalently as 
\begin{align}\label{eqn:kittelf}
&f(\theta, \phi) = A \pm \\ \nonumber
& \sqrt{B^2+C^2 \sin ^2(\theta ) \left[\cos^2 (\theta) + \cos^2 (\phi) \sin^2 ( \theta ) \sin^2 (\phi) \right]}.
\end{align}
The upper (positive) sign in \Eqns{eqn:kittel} and (\ref{eqn:kittelf}) corresponds to the heavy-hole band, while the lower (negative) sign generates the light-hole band. For $C=0$ both bands become spherical: otherwise, they are both warped to a degree increasing with the $C/B$ ratio.

\begin{figure*}[ht!]
	\begin{center}
 	\includegraphics[width=12.9cm]{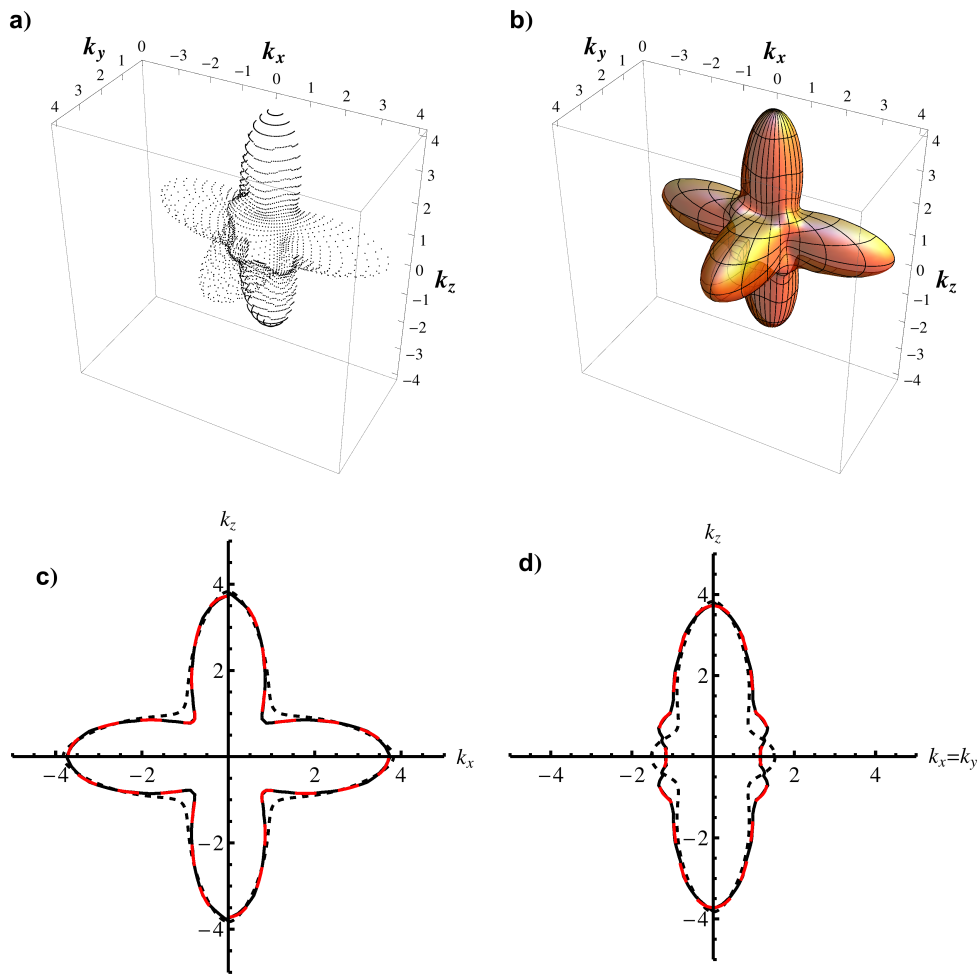}\\
      \caption{\label{kittelcompare} (Color) Angular effective mass surface $f(\theta,\phi)$ for the highest or heavy-hole valence band of Si at $\Gamma$, including spin-orbit interactions, as calculated from first principles using the $\textsc{Quantum Espresso}$ ($\textsc{QE}$) package.\cite{QE-2009} a) 2653 angular data points calculated using a quadratic fit with respect to $k_r$ of the energy dispersion for each angular direction. b) Spherical harmonics fit up to $l=20$ to the QE data in plot a). c)  Azimuthal $\phi = 0$ cross section of the effective mass surface. The black-dashed curve shows an interpolation from the data, the red-dashed curve represents a $l=20$ spherical harmonics fit, and the black-dotted curve shows a least-squares fit to the ``Kittel form'', \Eq{eqn:kittelf}, with optimal parameters $A = -4.20449$, $B=0.378191$, $C = 5.309$. d) Azimuthal $\phi = \pi/4$ cross section of the effective mass surface, with curves drawn as in c).}
	\end{center}
\end{figure*}

Our first-principles calculations suggest that the ``Kittel form'' represented by \Eqns{eqn:kittel} or (\ref{eqn:kittelf}) provides an overall good approximation only for the top heavy-hole band near $\Gamma$ in Si, corresponding to the upper (positive) sign, although even that is not uniformly accurate. In particular, it misses some notable features in comparison with the angular effective mass surfaces computed from first principles. In \Figref{kittelcompare}(a) we plot the angular effective mass surface $f(\theta,\phi)$ for the top heavy-hole valence band of Si at $\Gamma$, including spin-orbit interactions. That $f(\theta,\phi)$ is obtained by fitting \Eq{eqn:deff} to first-principles calculations, as previously described. We may also expand $f(\theta,\phi)$ in spherical harmonics as
\begin{align}\label{eqn:sphericalharmonics}
f(\theta, \phi) = \displaystyle\sum_{l=0}^{\infty} \, \displaystyle\sum_{m = -l}^{l} f_{lm} \, Y_{lm} (\theta, \phi).
\end{align}
An excellent fit with spherical harmonics up to $l=20$ to first-principles calculations is then plotted in \Figref{kittelcompare}(b). Cross sections of $f(\theta,\phi)$ in two azimuthal planes with $\phi = 0$ and $\phi = \pi/4$ are shown in Figs.~\ref{kittelcompare}(c) and (d), respectively. In particular, first principles calculations are compared with a least-squares fit to \Eq{eqn:kittelf} with optimal parameters $A = -4.20449$, $B = 0.378191$, $C = 5.309$.\footnote{By comparison, $A \simeq -4.1$, $B \simeq 1.6$, $C \simeq 3.3$ were obtained by fitting experimental data of cyclotron resonance in the original paper of Dresselhaus \textit{et al.}\cite{Dresselhaus}} That optimal fit shows the location and extent of inaccuracy inherent to \Eq{eqn:kittelf}, which occur primarily along diagonal coordinate directions. In any case, it is clear that none of the plots in \Figref{kittelcompare} can be sensibly related to, or derived from a sphere, having exclusively an $l=0$ component. But that is what the multi-dimensional quadratic expansion of \Eq{eqn:quad} would require for a cubic crystal, which has equivalent coordinate axes, implying equal effective masses and isotropic principal axes. The corresponding (mala fide) Hessian matrix would be proportional to the identity, excluding any band warping from the associated spherical quadratic expansion.

In Tab.\ \ref{tab:table1} we report warping parameters for heavy- and light-hole bands at $\Gamma$ in Si computed from first-principles calculations ($\textsc{QE}$) or from their fit with the ``Kittel form'', \Eq{eqn:kittel}. Referring to \Eq{eqn:COV}, $\sigma_w$ represents its numerator or RMS, while $\mu_w$ provides its denominator, or the angular average of the Hessian matrix trace. Both $\sigma_w$ and $\mu_w$ are dimensionless, having considered just $f(\theta,\phi)$ in the definition of the radial quadratic expansion, \Eq{eqn:deff}, and in the corresponding definitions of the Hessian matrix traces in \Eq{eqn:Traces}. In the last column of Tab.\ \ref{tab:table1}, we report the warping parameter $w$, represented by the ratio in \Eq{eqn:COV}. The most striking and robust results shown in Tab.\ \ref{tab:table1} are that the heavy-hole band is considerably warped and that is consistent with the ``Kittel form'' fit. Both such results were expected from those reported in \Figref{kittelcompare}. By contrast, the light-hole band appears to behave differently. It is not entirely clear why the ``Kittel form'' does not provide an approximation nearly as good for the light-hole band, compared to our first-principles calculations. Spurious effects originating from band non-parabolicity may be partly responsible for the poor fit of the light-hole band with the ``Kittel form.'' In any case, the small value of $w$ for the light-hole band is mainly the result of the large value of $\mu_w$ in the denominator of \Eq{eqn:COV}, which in turn derives from the greater dispersion or mean curvature of the light-hole band. This provides a good warning to consider both values of $w$ and $\sigma_w$ when assessing the magnitude of band warping in general.

\begin{table}
\caption{\label{tab:table1}Warping parameters for heavy- and light-hole bands at $\Gamma$ in Si computed from first-principles calculations ($\textsc{QE}$) or their fit with the ``Kittel form'', \Eq{eqn:kittel}.}
\begin{ruledtabular}
\begin{tabular}{lllll}
Band & Kittel/$\textsc{QE}$\footnote{Spherical harmonics fit up to $l=20$ in \Eq{eqn:sphericalharmonics}} & $\sigma_w$& $\mu_w$& $w$\footnote{\Eq{eqn:COV}}\\
\hline
\textrm{HH} & \textrm{Kittel} & 2.778 & -11.273 & -0.246 \\

\textrm{HH} & \textrm{\textsc{QE}} & 2.727 & -11.275 & -0.242\\

\textrm{LH} & \textrm{Kittel} & 2.778 & -39.180 & -0.071\\

\textrm{LH} & \textrm{\textsc{QE}} & 1.368 & -27.265 & -0.050\\
\end{tabular}
\end{ruledtabular}
\end{table}

Helmholz and Voon\cite{Helmholz2002} have compared the $\bd{k} \cdot \bd{p}$ form of \Eq{eqn:kittel} with other expressions mainly derived from tight-binding calculations for the valence bands of silicon. Differences are shown in their Figure 2(a) and Table 1, for example.  However, Helmholz and Voon \cite{Helmholz2002} did not have our angularly complete \Eq{eqn:deff}, and so they were able to consider and display only certain parts of the actual discrepancies among various results and fits obtained previously in the literature. 

\section{Transport coefficients}\label{trancoef}
\subsection{Theoretical Formulation}
We may now consider electrical and thermal electronic transport in anisotropic materials from a microscopic perspective, starting with Boltzmann's transport equation within the relaxation-time approximation.\cite{Ashcroft, GP, Marder} We introduce our angular effective mass formulation within a general formalism, as developed for example by Bies, Radtke, Ehrenreich, and Runge for anisotropic semiconductors.\cite{Bies2002} However, we will considerably broaden the types and features of their models and results. Most notably, our derivation, based on \Eq{eqn:deff}, generalizes conventional theories by including band-warping, multi-band, and anisotropy effects on electronic transport. Our derivation is substantively improved over the standard of Ref.~\onlinecite{Ashcroft}, and our analysis would further improve the interpretation of important elements recently discovered.\cite{ParkerPRL2013, Chen2013, Usui2010, Filippetti2012, Shirai2013}

In a material, heat generation is given by the absolute temperature $T$ times the rate at which the entropy of carriers changes. That produces the relation $\bd{j}^{Q} = T \bd{j}^{S}$ between the thermal current density $\bd{j}^{Q}$ and the entropy current density $\bd{j}^{S}$. We may then cast thermodynamic laws at constant volume, $T dS = dU - \mu d N$, in the current form
\begin{equation}\label{eqn:entropycurrent}
\bd{j}^{Q} = T \bd{j}^{S} = \bd{j}^{U} - \mu \bd{j}^{\textrm{N}},  
\end{equation}
where $\bd{j}^{\textrm{N}}$ is the number current density of carriers and $\bd{j}^{U}$ is their energy current density, while $\mu$ is the chemical potential.

We now consider specifically electronic transport in a solid and use basic expressions for the corresponding heat and entropy currents in terms of a non-equilibrium distribution function $g_n (\bd{k})$, namely
\begin{equation}
\bfrac{\bd{j}^{U}}{\bd{j}^{\textrm{N}}} = \displaystyle\sum_{n} \int \frac{\dd \bd{k}}{4 \pi^3} \bfrac{E_n(\bd{k})}{1} \bd{v}_n(\bd{k}) \, g_n(\bd{k}).
\end{equation}
By inserting these expressions into \Eq{eqn:entropycurrent} we derive the energy current
\begin{equation}\label{eqn:currentU}
\bd{j}^{U} = \displaystyle\sum_{n} \int \frac{\dd \bd{k}}{4 \pi^3} (E_n(\bd{k}) - \mu ) \, \bd{v}_n(\bd{k}) \, g_n(\bd{k}),
\end{equation}
and the electrical, i.e., electronic current 
\begin{equation}\label{eqn:currentj}
\bd{j} = - e \displaystyle\sum_{n} \int \frac{\dd \bd{k}}{4 \pi^3} \bd{v}_n(\bd{k}) \, g_n(\bd{k}),
\end{equation}
where $e = |e|$ denotes the absolute value of the charge of the electron, whereby $\bd{j} = -e \bd{j}^{\textrm{N}}$.

The relaxation-time approximation to Boltzmann's transport equation may be developed at various levels of sophistication, depending on effects that one may wish to capture in phenomena of various complexity.\cite{Ashcroft, GP, Marder, ParkerPRL2013, Chen2013} Some authors have considered in particular the possibility of anisotropic scattering in many-valley semiconductors.\cite{Bies2002, Herring1956, Ito1964} We follow that lead and begin by assuming a most general $\bd{k}$-dependent tensor form for the relaxation time associated with each band $n$, namely
\begin{equation}
\underline{\bm{\tau}}_{n} = \underline{\bm{\tau}}_{n} (\bd{k}).
\end{equation}
More specific restrictions and conditions will be subsequently introduced and discussed.

In order to proceed, one may further assume that the non-equilibrium distribution function $g_n (\bd{k})$ differs only slightly from a state of equilibrium. One may then expand $g_n (\bd{k})$ to first order as
\begin{align}\label{eqn:distributionfunction}
g_n(&\bd{k}) = g^{0}_n (\bd{k}) +\nonumber\\& \left( -\frac{\pd f_0}{\pd E} \right) \left[ \underline{\bm{\tau}}_n \cdot \bd{v}_n (\bd{k}) \right] \cdot \left[ - e \bm{\mathcal{E}} + \frac{E_n(\bd{k}) - \mu}{T} (-\bm{\nabla} T) \right],
\end{align}
where $g^0_n (\bd{k})$ is the equilibrium distribution and $f_0(E)$ is the Fermi-Dirac distribution function, implicitly related to each other as $f_0(E) = g^0_n (E(\bd{k}))$.\cite{Ashcroft, GP} In \Eq{eqn:distributionfunction}, $\bm{\mathcal{E}} = \bd{E} + \bm{\nabla} \mu/e$ is the electromotive force that drives the electronic current within Ohm's Law, while $\bm{\nabla} T$ is the temperature gradient that drives the heat current within Fourier's Law.

In order to discuss thermoelectric effects, we must further consider off-diagonal coupling between those two currents in a tensorial form appropriate for anisotropic materials, namely
\begin{equation}\label{eqn:onsager}
\left( \begin{array}{l} \bd{j} \\  \bd{j}^{Q} \end{array} \right) = \left( \begin{array}{ll} \underline{\textbf{\textsf{L}}}^{11} & \underline{\textbf{\textsf{L}}}^{12} \\ \underline{\textbf{\textsf{L}}}^{21} & \underline{\textbf{\textsf{L}}}^{22} \end{array}\right) \cdot \left( \begin{array}{c} \bm{\mathcal{E}} \\  \frac{- \bm{\nabla} T}{T} \end{array} \right).
\end{equation}

Up to this point, we should remark that \Eq{eqn:onsager} can be derived on most general grounds as a result of linear response theory.\cite{Callen} We have not yet made any use of, or any assumption about Onsager relations, which will be introduced later (cf. \Eq{eqn:onsagerrelations}). Therefore, no particular relation may be presumed between the off-diagonal kinetic coefficients $\underline{\textbf{\textsf{L}}}^{12}$ and $\underline{\textbf{\textsf{L}}}^{2 1}$. In fact, by setting either $\bm{\mathcal{E}} = 0$ or $\bm{\nabla} T = 0$ in \Eq{eqn:onsager}, one can readily discern that $\underline{\textbf{\textsf{L}}}^{1 2}$ and $\underline{\textbf{\textsf{L}}}^{2 1}$ have quite different physical interpretations, bearing no \textit{a priori} relation to one another. Nevertheless, if we substitute \Eq{eqn:distributionfunction} into \Eqns{eqn:currentU} and (\ref{eqn:currentj}), and recast the results in the form of \Eq{eqn:onsager}, we can demonstrate that $\underline{\textbf{\textsf{L}}}^{1 2}$ and $\underline{\textbf{\textsf{L}}}^{2 1}$ turn out to be equal. This result is bound to have major consequences for the development of our theory of transport properties in general and for the structure of the Seebeck coefficient or thermopower tensor, $\underline{\bm{S}}$, in particular. Thus far, we must then regard the $\underline{\textbf{\textsf{L}}}^{1 2} = \underline{\textbf{\textsf{L}}}^{2 1}$ equality as a particular consequence resulting from the combination of the microscopic semi-classical theory based on Boltzmann's transport equation with the relaxation-time approximation, having considered only small linear deviations from thermal equilibrium.\cite{Ashcroft, GP}

Based on such microscopic theory, all four kinetic coefficient tensors ($\underline{\textbf{\textsf{L}}}^{11}$, $\underline{\textbf{\textsf{L}}}^{12}$, $\underline{\textbf{\textsf{L}}}^{21}$, and $\underline{\textbf{\textsf{L}}}^{22}$) can be expressed in terms of three transport tensors,
\begin{align}
\underline{\pmb{\mathscr{L}}}^{(\alpha)} &= e^2 \displaystyle\sum_{n} \int \frac{\dd \bd{k}}{4 \pi^3} \left( -\frac{\pd f_0}{\pd E} \right) \nonumber\\  &\phantom{e^2 \displaystyle\sum_{n} \int} \times \bd{v}_n(\bd{k}) \underline{\bm{\tau}}_{n} \cdot \bd{v}_n(\bd{k}) (E_n (\bd{k}) - \mu)^{\alpha} , \label{eqn:boltzmanL}
\end{align}
as $\underline{\textbf{\textsf{L}}}^{11} = \underline{\pmb{\mathscr{L}}}^{(0)}$, $\underline{\textbf{\textsf{L}}}^{21} = \underline{\textbf{\textsf{L}}}^{12} = -(1/e) \underline{\pmb{\mathscr{L}}}^{(1)}$, and $\underline{\textbf{\textsf{L}}}^{22} = (1/e^2) \underline{\pmb{\mathscr{L}}}^{(2)}$. Notice that integrals involving the equilibrium distribution $g^0_n (\bd{k})$ in \Eq{eqn:distributionfunction} vanished from \Eq{eqn:boltzmanL} because of odd parity. With regard to our tensorial notations, we follow the standard dyadic and contraction conventions.\cite{BSL} 

So far, our microscopic derivation does not impose any symmetry condition on $\underline{\pmb{\mathscr{L}}}^{\alpha}$. At this point, however, we can and should further introduce Onsager reciprocity relations, requiring
\begin{equation}\label{eqn:onsagerrelations}
\underline{\textbf{\textsf{L}}}^{\mu \nu} = (\underline{\textbf{\textsf{L}}}^{\nu \mu})^{T}.
\end{equation}
Notice that these relations have an entirely different origin, based on independent assumptions of microscopic reversibility and that the decay of spontaneous fluctuations coincides with macroscopic flow processes in thermodynamics.\cite{Bies2002, Callen}

Now, if we combine Onsager relations, \Eq{eqn:onsagerrelations}, with our microscopic semiclassical results, following \Eq{eqn:boltzmanL}, we arrive at the following remarkable conclusions. Given that $\underline{\textbf{\textsf{L}}}^{11} = (\underline{\textbf{\textsf{L}}}^{11})^{T}$ on account of Onsager reciprocity, \Eq{eqn:onsagerrelations}, our microscopic theory tensor $\underline{\pmb{\mathscr{L}}}^{(0)} = \underline{\textbf{\textsf{L}}}^{11}$ must be symmetric. For the same reason, $\underline{\pmb{\mathscr{L}}}^{(2)}$ must also be symmetric. Moreover, combining Onsager reciprocity relation $\underline{\textbf{\textsf{L}}}^{12} = (\underline{\textbf{\textsf{L}}}^{21})^{T}$ with the condition $\underline{\textbf{\textsf{L}}}^{21} = \underline{\textbf{\textsf{L}}}^{12}$ derived independently from our microscopic semi-classical theory, we conclude that $\underline{\pmb{\mathscr{L}}}^{(1)}$ must be a symmetric tensor as well. This symmetry characteristic of all three $\underline{\pmb{\mathscr{L}}}^{(\alpha)}$ transport tensors has been previously noted in special cases.\cite{Ashcroft, Bies2002} It will have even greater consequences and more profound implications for our further considerations of band-warped energy dispersions, multi-band models, and anisotropy, although one must always keep in mind the general assumptions on which all that has been based.

\subsection{Two-Band Model of Transport}
We may apply these fundamental concepts to the basic case of a two-band semiconductor (with either a direct or an indirect band energy gap) and derive explicitly its transport formul\ae. We shall later generalize this basic analysis to multi-band cases, arriving at \Eq{eqn:Ks}. 

Based on our angular effective mass formalism, we represent a two-band structure with extrema at $\bd{k}_n$ points that may or may not coincide in the BZ as
\begin{equation}\label{eqn:def E}
E(\bd{k}) = \left\{ 
\begin{array}{ll}
E_c + \frac{\hbar^2 k_r^2}{2 \me}  f_c(\theta, \phi) & \textrm{if} \, E > E_c \\
E_v - \frac{\hbar^2 k_r^2}{2 \me}  f_v(\theta, \phi) & \textrm{if} \, E < E_v
\end{array}
\right. .
\end{equation}
In \Eqns{eqn:def E} we have assumed that the conduction band has a minimum at $\bd{k}_c$ and that the valence band has a maximum at $\bd{k}_v$, while there are no energy states nor energy dispersion inside the band gap. We have correspondingly reset $k_r^2 = ( \bd{k} - \bd{k}_n )^2 = k_x^2 + k_y^2 + k_z^2$, with $n = c,v$, respectively. Thus, the signs of both $f_c(\theta, \phi)$ and $f_v(\theta,\phi)$ are assumed to be positive everywhere.\footnote{That is not a fundamental requirement, however. For saddle points, $f(\theta, \phi)$ changes sign in different $\bd{k}$-regions, but $\bd{k}$-space integrations can be partitioned accordingly. Similar considerations later apply to \Eq{eqn:multi} and \Eq{eqn:Ks} and are further discussed in Appendix\ \ref{appendix:transformation}.} 
 
Using the definition of the electron mean or semi-classical velocity,\cite{Ashcroft, GP} and expressing that in polar coordinates,
\begin{equation}\label{eqn:vs}
\bd{v}_{n} = \frac{1}{\hbar} \frac{\pd E_n}{\pd \bd{k}} = \pm \frac{\hbar}{2 \me} \mathbb{J}^{-1} \left( \begin{array}{c} \frac{\pd E_n}{\pd k_r} \\ \frac{\pd E_n}{\pd \theta} \\  \frac{\pd E_n}{\pd \phi} \end{array} \right) = \pm \frac{\hbar k_r}{2 \me} \bd{\hat{v}}_{n},
\end{equation} 
we obtain the polar coordinate representation of the dimensionless vector in the velocity direction defined in \Eq{eqn:vs} as
\begin{align}\label{vstwo}
&\bd{\hat{v}}_{n} = \\ \nonumber 
&\left(\begin{array}{c} 2 \cos \phi \sin \theta f_{n}(\theta, \phi)+\cos \theta \cos \phi \frac{\pd f_{n}}{\pd \theta} -\csc \theta \sin \phi \frac{\pd f_{n}}{\pd \phi} \\ 2 \sin \theta \sin \phi f_{n}(\theta, \phi) + \cos \theta \sin \phi \frac{\pd f_{n}}{\pd \theta} + \cos \phi \csc \theta \frac{\pd f_{n}}{\pd \phi} \\ 2 \cos \theta f_{n}(\theta, \phi)- \sin \theta \frac{\pd f_{n}}{\pd \theta} \end{array}\right). 
\end{align}
That applies to the conduction band for $n=c$ and to the valence band for $n=v$, respectively. In order to derive \Eqns{eqn:vs} and (\ref{vstwo}), we used the Jacobian matrix $\mathbb{J}$ of Cartesian-to-polar coordinate transformation.

We should remain mindful that explicit consideration of any $f_{n}(\theta, \phi)$ requires the selection of a specific coordinate system. Therefore, we must also express the components $\left[ \underline{\bm{\tau}}_{n} \right]_{i,j}$ of the relaxation-time tensor in the same coordinate system. Matrix elements of transport tensors for the conduction band correspondingly become

\begin{align}
\left[ \underline{\pmb{\mathscr{L}}}_{c}^{(\alpha)} \right]_{i,j} &=  \nonumber\\ &\frac{ e^2 \, \hbar^2}{16 \pi^3 \me^2} \int \hat{v}_{c \, i} \sum_{q} \left[ \underline{\bm{\tau}}_{c} \right]_{j,q} \hat{v}_{c \, q} \nonumber\\& \times k_r^2 (E - \mu)^\alpha \left( -\frac{\pd f_0}{\pd E} \right) \dd k_x \dd k_y \dd k_z.
\end{align}

A similar expression applies for the valence band.

By relabeling $\hbar^2 k_i^2 / 2 \me \rightarrow k_i^2$ and changing variables to $(E,\theta, \phi)$, as detailed in Appendix\ \ref{appendix:transformation}, we obtain the expressions for the transport tensors comprising both bands as
\begin{widetext}
\begin{align}\label{eqn:boltzkijnew}
\left[ \underline{\pmb{\mathscr{L}}}^{(\alpha)} \right]_{i,j}&= \left[ \underline{\pmb{\mathscr{L}}}^{(\alpha)}_{c} + \underline{\pmb{\mathscr{L}}}^{(\alpha)}_{v} \right]_{i,j} =\nonumber\\ &\frac{e^2 \sqrt{\me}}{2^{3/2} \pi^3 \hbar^3} \left[ \int_{E_c}^{\infty} (E - \mu)^\alpha \left( -\frac{\pd f_0}{\pd E} \right) (E - E_c)^{3/2} C_{c \, i,j} \, \dd E \right. + \left. \int_{-\infty}^{E_v} (E - \mu)^\alpha \left( -\frac{\pd f_0}{\pd E} \right) (E_v - E)^{3/2} C_{v \, i,j} \, \dd E \right].
\end{align}
\end{widetext}
In \Eq{eqn:boltzkijnew} we have introduced for each band $n$ a constant tensor with matrix elements
\begin{align}\label{eqn:ctensors}
\left[ \underline{\bm{C}}_n \right]_{i,j} &= C_{n \, i,j} = \nonumber\\ &\sum_{q} \int_{0}^{2 \pi} \! \! \! \int_{0}^{\pi} \frac{\hat{v}_{n,i}(\theta,\phi) \left[ \underline{\bm{\tau}}_{n} \right]_{j,q} \hat{v}_{n,q}(\theta,\phi)}{2 |f_{n}(\theta,\phi)|^{5/2}} \sin \theta \, \dd \theta \dd \phi,
\end{align}
which will prove to be essential for the development of our analysis. Its definition in \Eq{eqn:ctensors} involves a particularly weighted average over the angular effective mass function $f_{n} (\theta, \phi)$ introduced in \Eq{eqn:deff}. Averaging over that angular function captures the essential effects of the electronic structure, including band warping, on all related transport coefficients. Using Onsager relations and a more elaborate derivation shown in Appendix\ \ref{appendixU}, we have further demonstrated that each $\underline{\bm{C}}_n$ must be a symmetric tensor.

It may also be worth noting that in \Eq{eqn:boltzkijnew}, as elsewhere, the negative of the derivative of the Fermi-Dirac distribution function
\begin{equation}\label{eqn:dfermi}
\left( \frac{-\pd f_0(E;\mu,\beta)}{\pd E} \right) = \frac{\beta e^{\beta (E-\mu)}}{(1+e^{\beta (E-\mu)})^2}
\end{equation}
is always positive, regardless of whether integration is performed for electron ($E > E_c$) or hole ($E< E_v$) states: cf.\ Ref.~\onlinecite{Wannier}, p.\ 290. Since it represents the probability density of occupation of states of any energy $E$ within $\dd E$, the Fermi-Dirac distribution function $f_0(E;\mu,\beta)$ does not intrinsically vanish inside the energy gap, where $E_v < E < E_c$, whatever the chemical potential $\mu$ and the temperature $T$ in $\beta = 1 / k_{\textrm{B}} T$ may be. It is instead the density of states that vanishes inside the energy gap of an intrinsic semiconductor, as modeled in \Eq{eqn:def E}. That is indeed the feature that permits splitting energy integrals over each band separately in \Eq{eqn:boltzkijnew}, following the change of variables to $(E,\theta, \phi)$ demonstrated in Appendix\ \ref{appendix:transformation}.

Up to this point, we have maintained a most general dependence on $(E,\theta,\phi)$ in the relaxation-time matrix elements $\left[ \underline{\bm{\tau}}_{n} \right]_{j,q}$ included in the integrand of \Eq{eqn:ctensors} that averages out to $C_{n\,i,j}$. A simplifying assumption that is commonly made\cite{Ashcroft} is that relaxation times may depend explicitly only on the energy $E$. In that case, the matrix elements $\left[ \underline{\bm{\tau}}_{n} \right]_{j,q}$ factorize out of the integrations in $\dd \theta \dd \phi$ in \Eq{eqn:ctensors}, while integration in $\dd E$ remains to be performed in \Eq{eqn:boltzkijnew}. Further consideration should be given to the anisotropic tensorial nature of $\underline{\bm{\tau}}_{n}$. To simplify our notations, we will gloss over both issues in the following main text of this paper. Namely, we will consider an isotropic relaxation-time constant $\tau_{n}$ for each band, as it is commonly assumed in standard references\cite{Ashcroft, GP, Marder} and computational studies.\cite{wannier90-new, Boltztrap} More properly, however, we do provide a complete discussion of such relevant issues and their consequent generalizations of our formalism in Appendix\ \ref{appendixU}.

We may thus simplify \Eq{eqn:boltzkijnew} as
\begin{align}\label{eqn:twobandL}
\left[ \underline{\pmb{\mathscr{L}}}^{(\alpha)} \right]_{i,j} =& \tau_c \, \mathcal{C}_{c \, i,j} K_\alpha (\beta (E_c - \mu), \beta) \nonumber\\ &+ (-1)^{\alpha} \tau_v \, \mathcal{C}_{v \, i,j} K_\alpha (\beta (\mu - E_v) , \beta),
\end{align}
where we define ``universal'' functions $K_\alpha$ as
\begin{equation}\label{eqn:universalKn}
K_\alpha (s , \beta) = \frac{e^2 \sqrt{\me}}{2^{3/2} \pi^3 \hbar^3 \beta^{\alpha + 3/2}} \int_{s}^{\infty} \frac{x^\alpha (x - s)^{3/2} e^x}{(1+e^x)^2} \, \dd x.
\end{equation}
As shown in Appendix\ \ref{appendixFD}, the integrals in $K_\alpha$ may also be expressed in terms of standard Fermi-Dirac integrals.\footnote{Consideration of at least energy-dependent isotropic relaxation times $\tau_{n}(E)$ would clearly alter the basic definition of \Eq{eqn:universalKn}, but only in an obvious manner that would not substantially modify our following results and discussions.} Most notably, we have introduced in \Eq{eqn:twobandL} a constant tensor with manifestly symmetric matrix elements
\begin{equation}\label{eqn:scriptC}
\mathcal{C}_{n \, i,j} = \left[ \underline{\bm{\mathcal{C}}}_{n} \right]_{i,j} =  \int_{0}^{2 \pi} \! \! \! \int_{0}^{\pi} \frac{\hat{v}_{n i}(\theta,\phi) \hat{v}_{n j}(\theta,\phi)}{2 |f_{n}(\theta,\phi)|^{5/2}} \sin \theta \, \dd \theta \dd \phi,
\end{equation}
corresponding to those of \Eq{eqn:ctensors}, but no longer dependent on relaxation times.

It may be instructive to first reduce these concepts to the simplest case of an ellipsoidal energy dispersion. Its angular effective mass function is readily derived as
\begin{equation}\label{eqn:quadpolar}
f(\theta, \phi) = \frac{\sin ^2(\theta) \left(m_x \sin ^2(\phi )+m_y \cos^2(\phi )\right)}{m_x m_y}+\frac{\cos^2(\theta)}{m_z}
\end{equation}
in the coordinate system associated with the ellipsoid principal axes, with corresponding diagonal effective masses $m_x, m_y, m_z$. The matrix elements of the $\underline{\bm{\mathcal{C}}}$ tensor in that intrinsic system take the simple form
\begin{equation}\label{eqn:ellipsoidalC}
\left[ \underline{\bm{\mathcal{C}}}_{\textrm{ellipsoid}} \right]_{i,j}= \left( \frac{8 \pi}{3} \frac{\sqrt{m_1 m_2 m_3}}{m_i \sqrt{\me}} \right) \delta_{i,j},
\end{equation}
where $m_{1,2,3}$ denote the effective masses in the ${x,y,z}$ principal directions, $\delta_{i,j}$ is the Kronecker symbol, and no summation convention is implied over the repeated $i$ index.  

The matrix that represents this ellipsoidal $\underline{\bm{\mathcal{C}}}$ tensor in any other coordinate system is readily determined by performing a corresponding similarity transformation with orthogonal matrices. We will need to consider such transformations when we perform multi-band calculations in which different $\underline{\bm{\mathcal{C}}}_{n}$ have different principal axes. 

\subsection{Multi-Band and Multi-Valley Models of Transport}\label{sec:Multi-Band}
We now consider generalizing \Eq{eqn:twobandL} to multi-valley and multi-band systems, having $N_b$ energy extrema $E_{n}$ at $\bd{k}_{n}$ points in the BZ. Many different combinations and arrangements are of course possible.

At one extreme, all $E_{n}$ may be different, although some $\bd{k}_{n}$ points may coincide. We may refer more specifically to those as multi-band configurations, involving direct or indirect energy gaps.

At the opposite extreme, all $E_{n}$ may be equal, while all $\bd{k}_{n}$ points may differ. We may refer more specifically to those as multi-valley configurations. For them, symmetry operations typically transform each $\bd{k}_{n}$ into other points in the set, forming a ``star'' of $n = 1, 2,\ldots, N_b$ equivalent points in the BZ.\cite{Bassani} The constant energy surfaces and expansions around each $\bd{k}_{n}$ also rotate and transform into one another according to symmetry operations of the lattice point group. We should mention in this context that we shall disregard within our current treatment of transport any inter-valley scattering or interference among multi-valley band structures.\cite{RescaRestaPRL} Such a relatively common,\cite{Bies2002} although not always warranted, assumption is adopted in order to partition and limit integrals within symmetry-restricted volumes in the BZ and thus resolve the resulting transport tensors $\underline{\pmb{\mathscr{L}}}^{(\alpha)}$ into sums over individual energy extrema near the Fermi energy.

We need to account for all kinds of mixed configurations between those two extremes, as they occur in actual materials or can be engineered to enhance transport and optical properties. 
 
After setting $k_r^2 = ( \bd{k} - \bd{k}_n )^2$, we may follow the basic idea of \Eq{eqn:deff} and express the energy dispersion, whether ellipsoidal or warped, as
\begin{equation}\label{eqn:multi}
E_{n} (\bd{k}) = E_{n} + \textrm{m}_n \frac{\hbar^2 k_r^2}{2 \me} f_{n} (\theta, \phi),
\end{equation}
where $\textrm{m}_n$ is either $+1$ if the band has a minimum, or $-1$ if it has a maximum, at the $\bd{k}_{n}$ point. The two-band tensors derived in \Eq{eqn:twobandL} can then be generalized to multi-valley and multi-band cases as 
\begin{equation}\label{eqn:Ks}
\left[ \underline{\pmb{\mathscr{L}}}^{(\alpha)} \right]_{i,j} = \sum_{n = 1}^{N_b} (\textrm{m}_n)^{\alpha} \tau_{n}  K_{\alpha} (\textrm{m}_n \beta (E_{n} - \mu), \beta) \, \mathcal{C}_{n \, i,j}.
\end{equation}

In other cases that include saddle points, \Eq{eqn:multi} and \Eq{eqn:Ks} can be maintained in form, but bands and corresponding integrations must be split in $\bd{k}$-space regions with alternating $\textrm{m}_n$ values.\footnote{Similar considerations have already been made with regard to \Eqns{eqn:def E} in a corresponding footnote and are further mentioned in Appendix\ \ref{appendix:transformation}.}

For general multi-valley and multi-band models, we may now determine transport properties by extending standard procedures\cite{Bies2002, Ashcroft, Marder, GP} on the basis of our formalism. Primarily, the electrical conductivity tensor, $\underline{\bm{\sigma}}$, the Seebeck coefficient or thermopower tensor, $\underline{\bm{S}}$, and the electronic thermal conductivity tensor, $\underline{\bm{\kappa}}$, can be determined as   
\begin{subequations}
\label{eqn:transportdefs}
\begin{align}
\underline{\bm{\sigma}} &= \underline{\pmb{\mathscr{L}}}^{(0)}, \\\label{eqn:transportdefsb}
\underline{\bm{S}} &= \frac{-1}{T e} \left( \underline{\pmb{\mathscr{L}}}^{(0)} \right)^{-1} \cdot \underline{\pmb{\mathscr{L}}}^{(1)},\\
\underline{\bm{\kappa}} &= \frac{1}{T e^2} \left( \underline{\pmb{\mathscr{L}}}^{(2)} - \underline{\pmb{\mathscr{L}}}^{(1)} \cdot \left(\underline{\pmb{\mathscr{L}}}^{(0)} \right)^{-1} \cdot \underline{\pmb{\mathscr{L}}}^{(1)}\right).
\end{align}
\end{subequations}
Those tensors in turn contribute to thermoelectric figures of merit, the most common of which is $ZT$, tensorially expressed as $\underline{\bm{S}}^{T}\cdot \underline{\bm{\sigma}} \cdot\underline{\bm{S}} \cdot\underline{\bm{\kappa}}^{-1} T$. 

A special case with a particularly interesting and somewhat surprising result has already been explored by Bies \textit{et al.}\cite{Bies2002} That is the case of a multi-valley energy dispersion with degenerate energy minima $E_0$ placed in $N_b$ different valleys. In that case, \Eq{eqn:Ks} reduces to
\begin{align}\label{eqn:onebandmultivalley}
\left[ \underline{\pmb{\mathscr{L}}}^{(\alpha)}\right]_{i,j} &= \sum_{n = 1}^{N_b} \tau_{n} \, \mathcal{C}_{n \, i,j} K_\alpha (\beta (E_{0} - \mu), \beta) \nonumber\\ &= K_\alpha (\beta (E_{0} - \mu), \beta) \, \mathbb{B},
\end{align}
where
\begin{equation}\label{eqn:biesequation}
\mathbb{B}_{i,j} = \sum_{n = 1}^{N_b} \tau_{n} \, \mathcal{C}_{n \, i,j}
\end{equation}
is a constant and symmetric matrix, being the sum of constant and manifestly symmetric matrices, according to \Eq{eqn:scriptC}. Since the matrix elements of both $\underline{\pmb{\mathscr{L}}}^{(0)}$ and $\underline{\pmb{\mathscr{L}}}^{(1)}$ are then proportional to $\mathbb{B}$, the thermopower tensor, $\underline{\bm{S}}$, is merely proportional to the identity tensor, as a result of \Eq{eqn:transportdefsb}. Notice, however, that if the condition that all energy minima are degenerate is relaxed, such a nice factorization as in \Eqns{eqn:onebandmultivalley} and (\ref{eqn:biesequation}) can no longer occur and $\underline{\bm{S}}$ ceases to be isotropic. In fact, $\underline{\bm{S}}$ may not even be a symmetric tensor, when only \Eq{eqn:Ks} generally applies. We will provide a basic illustration of that anisotropy in Sec.\ \ref{sec:equivalentellipsoid}.

Let us then reiterate that if there are at least \emph{two} different and non-degenerate bands contributing to transport, since each $\underline{\pmb{\mathscr{L}}}^{(\alpha)}$ is a sum of different tensors in \Eq{eqn:Ks}, $(\underline{\pmb{\mathscr{L}}}^{(0)})^{-1} \cdot \underline{\pmb{\mathscr{L}}}^{(1)}$ is not generally a tensor proportional to the identity tensor, and it may not even be a symmetric tensor. It can be shown, however, that it still has real eigenvalues, although not necessarily orthogonal eigenvectors.

On the other hand, if there is just a \emph{single} band, the thermopower is proportional to the identity tensor and it has the following explicit expression:
\begin{align}
\underline{\bm{S}} &= \frac{-1}{T e} \left( (\underline{\pmb{\mathscr{L}}}^{(0)})^{-1} \cdot \underline{\pmb{\mathscr{L}}}^{(1)} \right) \nonumber\\ &= \frac{-1}{T e} \left( \frac{1}{K_0(\beta (E_0 - \mu), \beta)} \mathbb{B}^{-1} \cdot \mathbb{B} K_1(\beta (E_0 - \mu), \beta) \right) \nonumber\\
&= \frac{-1}{T e} \left( \frac{K_1(\beta (E_0) - \mu, \beta)}{K_0(\beta (E_0) - \mu, \beta)} \right) \mathbbm{1}.
\end{align}

Exactly the same expression holds for $N_b$ multi-valley or multi-band minima with the same degenerate energy $E_0$, independently of $N_b$ and of all the $\tau_{n}$ relaxation times, stemming from \Eqns{eqn:onebandmultivalley} and (\ref{eqn:biesequation}).

\section{equivalent-transport Ellipsoid and Anisotropy}\label{sec:equivalentellipsoid}
The symmetry of the $\underline{\bm{C}}$ tensor that we discussed in the preceding subsection (V-C) and demonstrated more generally in Appendix\ \ref{appendixU} has major consequences for transport. In particular, the process of integrating out all angular information in the calculation of $\underline{\bm{C}}$ may lead to the same value regardless of the form of the original band structure, whether that was spherical (having $f_{n}(\theta,\phi) = \textrm{const}$) or ellipsoidal (as in \Eq{eqn:quadpolar}) or warped (as in \Eq{eqn:kittelf} or most generally) at a critical point.

Since any real symmetric second-rank tensor originates a quadratic form, we have a prescription for generating an \emph{equivalent-transport ellipsoid} that can produce the same transport coefficients as any one band that may actually be warped. Notice that this is an entirely different matter than that of trying to approximate with an ellipsoid a warped band from the beginning by means of any fitting procedure, which we have already shown to be generally impossible. On the contrary, the procedure to determine an equivalent-transport ellipsoid for each band is always possible, following these three simple steps. (1) Given any warped band, first compute the integrals that provide the six independent matrix elements of $\underline{\bm{C}}$ or $\underline{\bm{\mathcal{C}}}$ tensors corresponding to \Eq{eqn:ctensors} or to \Eq{eqn:scriptC}. (2) Associate the latter with the rotated expression\footnote{Here $\mathbb{A}$ is an orthogonal matrix whose columns are unit vectors pointing in the direction of the new coordinate axes.} of \Eq{eqn:ellipsoidalC}, i.e., $\mathbb{A} [\underline{\bm{\mathcal{C}}}_{\textrm{ellipsoid}} ] \mathbb{A}^{T}$, where equivalent-transport masses and Euler angles remain to be determined. (3) Finally, solve six equations for six unknowns, determining three equivalent-transport masses and three Euler angles of rotation to a coordinate system of effective principal axes. This procedure clearly defines an equivalent-transport ellipsoid for the original band, no matter how warped. 

To provide an example, let us compute a (valid) equivalent-transport ellipsoid for the heavy-hole band in silicon and compare that with the (invalid) spherical approximation to that band, warped according to \Eq{eqn:kittel}. For that ``Kittel form'' of band warping and an isotropic relaxation time, the resulting $\underline{\bm{\mathcal{C}}}$ tensor must be proportional to the identity, because the lattice has cubic symmetry. By fitting first-principles calculations to \Eq{eqn:kittelf}, we derived earlier in Sec.\ \ref{sec:silicon} the optimal parameter values of $A = -4.20449$, $B = 0.378191$, and $C = 5.309$. The corresponding expression for the matrix elements of the $\underline{\bm{\mathcal{C}}}$ tensor in \Eq{eqn:scriptC} is $\left[\underline{\bm{\mathcal{C}}} \right]_{i,j} = 9.0102 \, \delta_{i,j}$ for the heavy-hole band. The corresponding result for the light-hole band is $\left[ \underline{\bm{\mathcal{C}}} \right]_{i,j} = 3.4853 \, \delta_{i,j}$. Associating those with corresponding equivalent-transport spherical bands, having $\underline{\bm{\mathcal{C}}} = (8 \pi/3) \sqrt{m_{\text{equiv}}}\, \openone$, results in equivalent-transport masses $m_{\text{equiv} \, hh} = 1.1567 \me$ for the heavy hole and $m_{\text{equiv} \, lh} = 0.17308 \me$ for the light hole.

On the other hand, if we unjustifiably try to fit from the start the band-warped surface to a sphere, e.g., by considering just the $l=0$ contribution in \Eq{eqn:sphericalharmonics} for the spherical harmonics expansion of the warped band, we come up with a (mala fide) spherical-average mass of $m_{(l=0) \, hh} = 0.53224 \, \me$ for the heavy hole and $m_{(l=0) \, lh} = 0.15314 \, \me$ for the light hole. Thus, even though a band-warped surface may always be replaced, insofar as transport properties are concerned, with an equivalent ellipsoidal band, that does not mean that band warping is irrelevant by any means. The absolute difference between the valid equivalent-transport mass $m_{\textrm{equiv} \, hh} = 1.1567 \me$ and the (mala fide) spherical-average mass of $m_{(l=0) \, hh} = 0.53224 \, \me$ provides compelling evidence of a major band-warping effect on transport properties, corresponding to the large band-warping parameter $w = -0.2465$ that we have previously derived in Sec.\ \ref{sec:silicon} for the heavy-hole band in silicon. 

Since heavy- and light-hole bands become degenerate at $\Gamma$, we may also add their $\underline{\bm{\mathcal{C}}}$ tensors, assuming that degenerate bands at a critical point also share a constant common relaxation time.\footnote{Otherwise, we must recall \Eq{eqn:biesequation} and modify the corresponding expressions provided in this paragraph and in Sec.\ \ref{undewar} for the combined equivalent-transport $\bm{\mathcal{C}}$ tensors and masses by weighting them with different relaxation times.  We skip such relatively straightforward generalizations in this context for the sake of simplicity.} In this degenerate case, a single equivalent-transport ellipsoid, which actually reduces to a sphere because of the cubic symmetry, produces the same transport properties of both heavy- and light-hole bands combined. We may then describe the corresponding transport as produced by a single carrier with a combined equivalent-transport mass $m_{\textrm{equiv} \, hh-lh} = \left( \frac{3}{8 \pi} (\mathcal{C}_{hh}+\mathcal{C}_{lh})\right)^2$. That is a special case of a more general multi-valley or degenerate multi-band $m_{\textrm{equiv combined}} = \left( \sum_j \sqrt{m_{\textrm{equiv} \, j}}\right)^2$, provided that all valleys or degenerate bands are of the same type, i.e., either conduction-like or valence-like. In our case, the resulting mass $m_{\text{equiv} \, hh-lh} = 2.22 \me$ of the combined carrier is related non-linearly to both $m_{\text{equiv} \, hh} = 1.1567 \me$ and $m_{\text{equiv} \, lh} = 0.17308 \me$ of the two previous individual equivalent-transport spheres, but it resembles neither.

\begin{figure}[ht!]
	\begin{center}
 	\includegraphics[width=5.5cm]{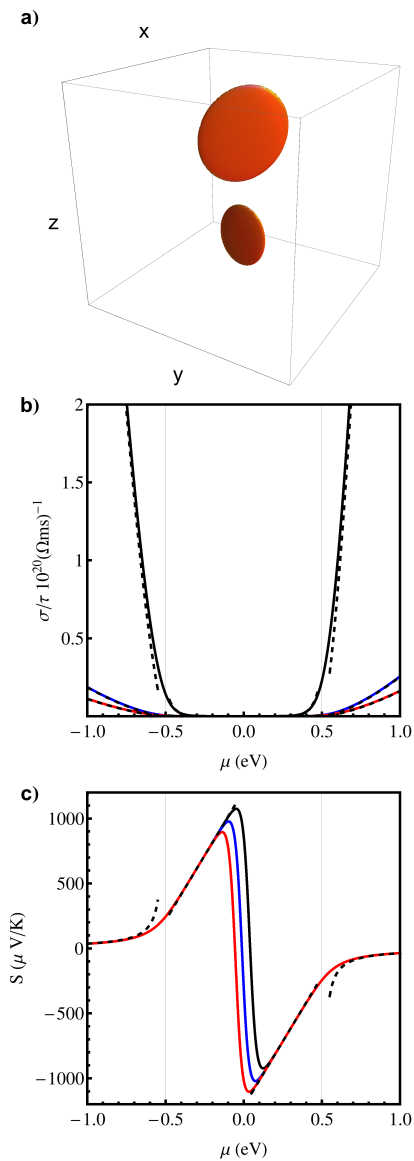}\\
      \caption{\label{fig:twoband2} (Color) Conductivity and thermopower tensor eigenvalues for a model with two highly anisotropic ellipsoidal bands. The two ellipsoidal bands have masses $m_{v,x} = 0.1, m_{v,y} = 3,m_{v,z} = 5$, and $m_{c,x'} = 0.2, m_{c,y'} = 7,m_{c,z'} = 11$, respectively. The principal axes of the conduction band are rotated with respect those of the valence band through Euler angles $\Phi = \pi/7$, $\Theta = \pi/6$, and $\Psi = 0$. Two constant-energy surfaces for the two bands are shown in panel (a). The valence-band maximum is set at -0.5 eV and the conduction-band minimum is set at +0.5 eV. Panels (b) and (c) show three curves each, representing the three eigenvalues of the conductivity tensor, normalized by the relaxation time, $\tau$, and of the thermopower tensor, as functions of the chemical potential, $\mu$. The absolute temperature is set at $T=500\,$K. One-band asymptotic expansions, as derived in Appendix\ \ref{appendixFD}, are represented by black dashed curves.}
	\end{center}
\end{figure}

Greater richness of effects on transport properties arise from consideration of equivalent-transport ellipsoids as a result of their \textit{anisotropy}, which we expect to find in non-cubic materials, or as a result of perturbations, such as strain or doping, or fabrication, as in heterostructures, superlattices, nano-structures, or nano-wires. Here we may provide just a sketch of the basic idea by considering a two-band model with substantially flattened equivalent-transport ellipsoids, shown in \Figref{fig:twoband2}(a).  As expected, the major relative reduction of one effective mass in a principal direction for each band ellipsoid produces corresponding enhancements and reductions in the eigenvalues of the conductivity tensor. The dependence on the chemical potential of the eigenvalues of the thermopower tensor is more subtle, but perfectly consistent with our discussion in the last two paragraphs of the preceding subsection V-C. Namely, $\underline{\bm{S}}$ becomes essentially isotropic within each energy band, where the contribution to transport from the other band becomes negligible. However, when the chemical potential lies between the energies of the two bands and both bands contribute to transport, the thermopower tensor is not even symmetric and its eigenvalues split, although they remain real. Since the principal axes of the the two equivalent-transport ellipsoids are rotated relative to each other, the eigenvectors of $\underline{\bm{S}}$ are not even orthogonal when both bands contribute to transport, whereas they approach the principal axes of either ellipsoidal band when the chemical potential falls into the energy of that band and away from the energy of the other band.

Further understanding may be gained from one-band expressions developed via asymptotic expansion in Appendix\ \ref{appendixFD}. Using those expansions, we can provide analytic expressions for $\underline{\bm{\sigma}}$ and $\underline{\bm{S}}$ when only one band is essentially contributing to the whole transport, either in the insulator or in the metallic limit, represented by \Eqns{Appendixeqn:transporteqnsinsulating} and (\ref{Appendixeqn:transporteqnsmetal}), respectively. These analytic expressions are represented by black dashed curves in \Figref{fig:twoband2}(b) and (c). Each asymptotic form is plotted only within its range of applicability, namely, away from band edges, on the order of $k_{\textrm{B}} T$. In \Figref{fig:twoband2}(b), the analytic asymptotic expressions indeed predict the exact behavior of \emph{each} branch of the electrical conductivity eigenvalues, away from band edges on the order of $k_{\textrm{B}} T$. In \Figref{fig:twoband2}(c), the analytic asymptotic expressions predict that the $\underline{\bm{S}}$ eigenvalues become independent of mass values and essentially equal, isotropically. However, sufficiently close to the middle of the energy gap, around $\mu = 0$ eV, the exact eigenvalues of $\underline{\bm{S}}$ differ considerably, because carriers from both bands contribute jointly to transport.

At the conclusion of Appendix\ \ref{appendixFD}, we also provide asymptotic expressions for the eigenvalues of $\underline{\bm{\sigma}}$ and $\underline{\bm{S}}$ in terms of the carrier density, $n$. Those expressions can be further used to broadly generalize implications of the Cutler-Mott formula\cite{Cutler, ParkerPRL2013, Chen2013, Filippetti2012, Snyder2008complex} to warped, anisotropic, and multi-band cases in both the insulator and metallic limits. 

Notwithstanding the major effects of anisotropy that are clearly illustrated by differing eigenvalues of the conductivity and thermopower tensors shown in \Figref{fig:twoband2}, it is worth recalling that scalar effective electrical and thermal conductivities, thermopower and $ZT$ figure of merit may still be sensibly defined and extracted, albeit laboriously, from all components of their fully tensorial counterparts, at least in anisotropic materials rectangularly cut and placed in appropriate configurations of applied generalized forces and response fluxes.\cite{Bies2002}  

\section{Effects of band warping on transport coefficients}\label{undewar}
Let us now illustrate the influence that band warping may have on electronic transport properties with some basic examples, which further emphasize the value of the equivalent transport formalism developed in Sec.\ \ref{sec:equivalentellipsoid}. On the other hand, we wish to leave out for the purpose of this discussion any consequence of \textit{anisotropy}, which has already been demonstrated in Sec.\ \ref{sec:equivalentellipsoid}. Thus we combine spherical bands with \textit{warped} bands that have at least a \textit{cubic} symmetry. For the latter, we further assume the prototypical ``Kittel form'' of \Eq{eqn:kittel}, where band warping is quantitatively tied to numerical values of the $A$, $B$, $C$ parameters. Those can be varied in computational experiments and models to evaluate corresponding effects on conductivity and thermopower tensors.

Although band warping is typically associated with band degeneracy at critical points, we need not focus on that aspect in this context. In the first of two illustrative examples, we will thus focus on a single (heavy-hole) band, disregarding any possible consideration of a companion (light-hole) band. In fact, even energy bands that are non-degenerate at critical points can be warped.\cite{ParkerPRL2013, Chen2013} In the second of our two illustrative examples, however, we will consider two degenerate bands. For comparison, we will fix the temperature at $T=500\,$K in both examples.

\begin{figure*}[ht!]
\begin{center}
  \includegraphics[width=12.9cm]{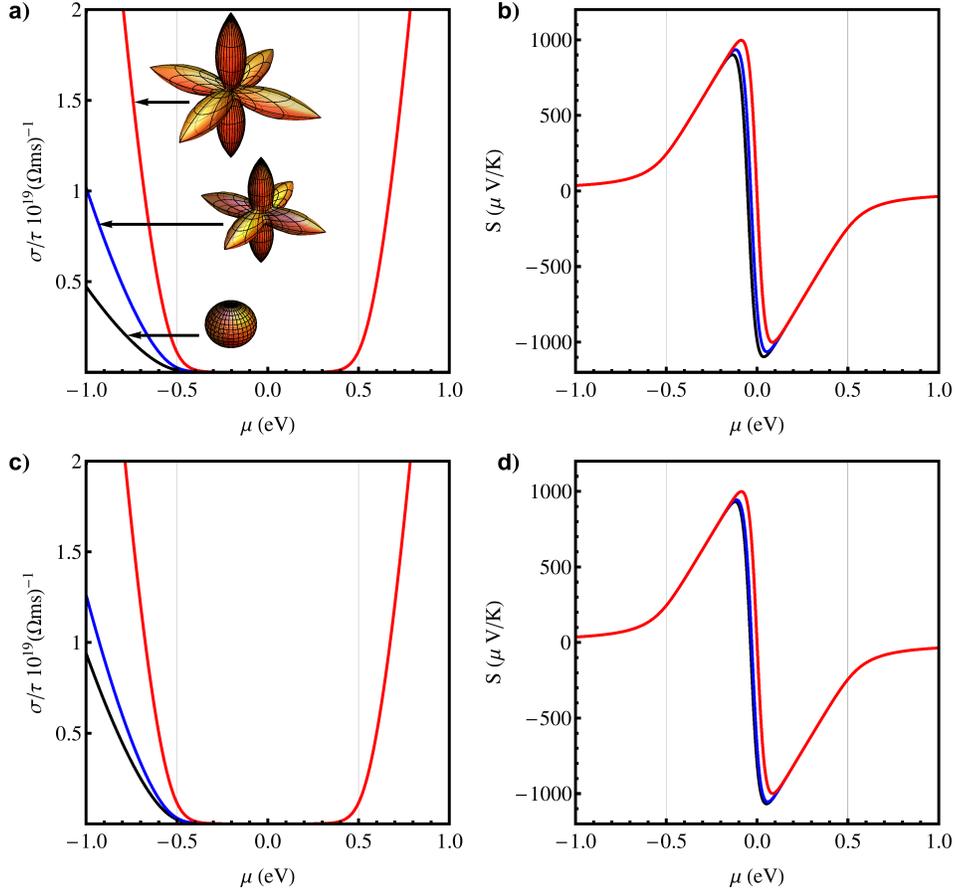}\\
      \caption{\label{twomodels} (Color) Isotropic conductivity and Seebeck coefficient for two models, comprising either one (panels (a) and (b)) or two (panels (c) and (d)) warped valence bands of the ``Kittel form'', at three different levels of warping. The maximum of all valence bands is set at -0.5 eV. Both models further comprise a single spherical conduction band with effective mass $m = \me$ and its minimum set at +0.5 eV. Each panel shows three curves, representing a negligible (black), medium (blue), and high (red) level of warping for the valence bands. In panel (a), the insets pointing at the valence-band curves at their three different levels of warping depict their corresponding angular effective mass surfaces. The isotropic conductivity, normalized by the same relaxation time for all bands, $\tau$, and the Seebeck coefficient are shown as functions of the chemical potential, $\mu$. The absolute temperature is set at $T=500\,$K in all cases.}
\end{center}
\end{figure*}

\begin{table*}
\caption{\label{tab:table2}Kittel, warping, and mass parameters corresponding to the plots of \Figref{twomodels}.  Zero, Mid and High warping parameters $w$ for the ``Kittel form'' of \Eq{eqn:kittelf}, where the B parameter may be taken as 1. For each pair of the three $w$-value rows, the first or heavy-hole (HH) line corresponds to our first model alone, while the second or light-hole (LH) line is added for our second model. }
\begin{ruledtabular}
\begin{tabular}{ccccccccc}
Band & Kittel & Kittel & Warping ($w$) & Spherical    & $m_{(l=0)}$             &    $\mathcal{C}$  & Equivalent & Double-Band \\
     & A/B    & C/B    &               & Approx.\footnote{$f_{00} Y_{00}$}             
                                                      & (in units of $\me$)\footnote{$-1/(f_{00} Y_{00})$}       
                                                                                &                   & Mass       & Combined Mass \\
     &        &        &               &  $(l=0)$ 
                                                      &  
                                                                                &                   & (in units of $\me$)\footnote{$(3 \mathcal{C}/ 8 \pi)^2$}                                                                                                               & (in units of $\me$)\footnote{$(3 (\mathcal{C}_1+\mathcal{C}_2)/ 8 \pi)^2$}\\
     &        &        &               &  in \Eq{eqn:sphericalharmonics}            
                                                      &                         &                   &            &               \\
\hline
\hline
\textrm{Zero $w$ HH} & -94.39 & 0 & $-4.64 \times 10^{-9}$ & -93.39 & 0.01071 & 0.8669 & 0.01071  &  \multicolumn{1}{c}{\multirow{2}{*}{0.0428}}\\
\textrm{Zero $w$ LH} &  &  & $-4.639 \times 10^{-9}$ & -95.39 & 0.01048 & 0.8669 & 0.01071  & \\
\hline
\textrm{Mid $w$ HH} & -256.5 & 380.0 & -0.3684 & -92.62 & 0.01080 & 1.869 & 0.04976  & \multicolumn{1}{c}{\multirow{2}{*}{0.0764}}\\
\textrm{Mid $w$ LH} &  &  & -0.08117 & -420.4 & 0.002379 & 0.4466 & 0.002840  & \\
\hline
\textrm{High $w$ HH} & -351.5 & 600.0 & -0.5808 &-92.77 & 0.01078 & 7.940 & 0.8983  & \multicolumn{1}{c}{\multirow{2}{*}{0.986}}\\
\textrm{High $w$ LH} &  &  & -0.08829 & -610.3 & 0.001639 & 0.3779 & 0.002030  & \\
\end{tabular}
\end{ruledtabular}
\end{table*}

Our first model represents the simplest (non-trivial) combination: one warped valence band and one spherical conduction band separated by an energy gap $E_g = 1$ eV. We also assume that the spherical conduction band has a free-electron effective mass $m = \me$ and it is thus represented by the $\underline{\bm{\mathcal{C}}} = (8 \pi/3) \, \openone$ tensor.  For the valence band, we begin by assuming a single warped band with the heavy-hole ``Kittel form'' in the case of zero, mid, and high warping represented with parameters $w = -4.64 \times 10^{-9}$, $-0.36843$ and $-0.5808$, respectively. By comparison, we may recall that the $w$ value for the heavy-hole band in Si is about -0.25: see Tab.\ \ref{tab:table1}.

In \Figref{twomodels} we plot the effects of band warping on the conductivity and thermopower. Both are represented by isotropic tensors proportional to the identity, and they are functions of the chemical potential, which is a variable controlled in theory by contact with a $T-\mu$ reservoir, or in practice by chemical doping. In this (grand-canonical ensemble) representation, the electronic conductivity is approximately proportional to the square root of an appropriate effective mass, while it does not depend explicitly on the density of carriers (as it does instead in the familiar canonical-ensemble representation). In the same (grand-canonical ensemble) representation, the Seebeck coefficient or thermopower depends only weakly on the chemical potential, at least deeply within each band.\footnote{For guidance, see the asymptotic expansions derived in Appendix\ \ref{appendixFD} in the metallic limit, i.e., \Eqns{Appendixeqn:transporteqnsmetal}. } This first example, illustrated in panels (a) and (b) of \Figref{twomodels}, demonstrates that band warping can substantially increase the electrical conductivity, while it has only a small effect on the Seebeck coefficient, except around the middle of the gap.

Numerical results can be more readily understood in terms of the equivalent-transport ellipsoid concept and the corresponding equivalent effective mass, $m_{\textrm{equiv}}$. In fact, the introduction of band warping simply amounts in this cubic symmetry case to replacing warped bands by equivalent-transport spheres with masses given by 0.01071 $\me$, 0.04976 $\me$, and 0.8983 $\me$, for the zero, mid, and high warping cases, as reported in Tab.\ \ref{tab:table2}. Our calculations show that $m_{\textrm{equiv}}$ for the heavy-hole type of valence band increases with band warping in Tab.\ \ref{tab:table2}, which in turn correctly results in a higher electronic conductivity in \Figref{twomodels}(a) at negative values of $\mu$, according to our grand-canonical ensemble representation: cf.\ \Eqn{Appendixeqn:transporteqnsmetal}(a) in Appendix\ \ref{appendixFD}. By contrast, since the conduction band is not warped, the conductivity is hardly affected by warping in \Figref{twomodels}(a) at positive values of $\mu$ because $\sigma$ is mainly determined by the spherical conduction band for {\it n}-type doping. 

It is also important to appreciate why the mass $m_{(l=0)}$ of the spherical approximation does not correlate with the increase in electrical conductivity caused by increase in band warping at negative values of $\mu$. In Tab.\ \ref{tab:table2}, we have arranged the three warping cases to have very similar $l=0$ masses, but very different $w$ values. For zero warping, represented by the black curve in \Figref{twomodels}(a), $C/B = 0$ implies that $f(\theta, \phi) = A + 1 = -93.39$ is spherical and constant. Hence $m_{(l=0)} = 0.01071$ $\me$, which is the inverse of $A+1 = -93.39$, must coincide with $m_{\text{equiv}}$, as confirmed in the first row of Tab.\ \ref{tab:table2}. On the other hand, for the mid and high warping cases, represented by the blue and red curves in \Figref{twomodels}(a), the $A/B$ and $C/B$ ratios are chosen to produce essentially the same $m_{(l=0)}$ as in the zero-warping case, but their $m_{\text{equiv}}$ are about four and eighty times larger than $m_{(l=0)}$, respectively. Evidently, the spherical approximation cannot account for any conductivity increase when band warping increases. In fact, we can quantitatively understand those conductivity increases. Recall that the conductivity equals $\underline{\pmb{\mathscr{L}}}^{(0)}$, that $\underline{\pmb{\mathscr{L}}}^{(0)}$ is proportional to $\underline{\bm{\mathcal{C}}}$, and that $\underline{\bm{\mathcal{C}}}$ is proportional to the square root of $m_{\text{equiv}}$, for cubic symmetry. In \Figref{twomodels}(a), the conductivity for the mid-warping case is approximately twice the zero-warping conductivity at negative values of $\mu$. This is quite consistent with the fact that $m_{\text{equiv}}$ for the mid-warping case is approximately four times larger than $m_{\text{equiv}}$ at zero warping. Further conductivity increase for the high-warping case can be correspondingly understood.

Our second model consists of two warped valence bands degenerate at their maximum $E_0 = -0.5$ eV, thus simultaneously contributing to transport, plus one spherical conduction band set as in the previous model at $E_c = +0.5$ eV. For these two valence bands, we use both the heavy-hole and the light-hole expressions of the ``Kittel form'', having upper (positive) and lower (negative) signs in \Eq{eqn:kittelf}. Due to their degeneracy at the critical point, warping of one valence band affects warping of the other band. We consider the same values of $A/B$ and $C/B$ ratios as in our first model.  For the heavy-hole bands, those ratios thus produce the same values of the warping parameter, i.e., $w = -4.64 \times 10^{-9}$, $-0.36843$ and $-0.5808$, as in our first model. For the light-hole bands, however, the corresponding levels of warping turn out to be substantially reduced, having values of $w = -4.64 \times 10^{-9}$, $-0.0811$, and $-0.0882$, respectively.  Such smaller values of $w$ for light-hole bands are mainly the result of larger values of $\mu_{w}$ in the denominator of \Eq{eqn:COV}, that in turn are a consequence of greater dispersion or mean curvature for light-hole bands. We have already noticed that effect in Tab.\ \ref{tab:table1}, when reporting parameters for heavy-hole and light hole bands at $\Gamma$ in Si computed from first-principles calculations ($\textsc{QE}$) or from their fit with the ``Kittel form".  

The introduction of an additional valence band contributes further to the increase of the conductivity in our second model compared to our first model. Indeed, notice in panels (a) and (c) of \Figref{twomodels} that, for zero warping (black lines), the electrical conductivity essentially doubles because of the inclusion of a second, virtually identical band, their equivalent masses being the same. This is because their combined equivalent mass is four times greater than each of the individual masses, effectively doubling the conductivity. When warping increases, however, the two bands strongly interact. Their equivalent effective masses, reported in Tab.\ \ref{tab:table2}, reflect such a coupling by increasing substantially for the heavy hole, while decreasing drastically for the light hole.  Nonetheless, the conductivity is accounted for by their combined equivalent mass, reported in the last column in Tab.\ \ref{tab:table2}.

The quantitative behavior shown in panel (c) of \Figref{twomodels} shows that increasing band warping  again causes the electrical conductivity to increase, even a bit more than in our first model, shown in panel (a) of \Figref{twomodels}, because our second model has an additional light-hole band. Still, the general trends observed in panels (c) and (d) of \Figref{twomodels} are overall consistent with the increase in $m_{\text{equiv} \, hh-lh}$ for the single equivalent-transport ellipsoid that combines warping of both valence bands.

The emergence of the equivalent-transport ellipsoid concept has profound implications. On the one hand, it prevents direct observation of band warping in transport measurements. Since the basic $\underline{\bm{\mathcal{C}}}$ tensor that determines an equivalent-transport ellipsoid is defined in \Eq{eqn:scriptC} through an angular integration average, there is a many-to-one functional relation between angular effective mass functions $f(\theta,\phi)$ and the same equivalent ellipsoid that represents them for transport purposes. For instance, we can describe two degenerate valence bands active in transport, as in our second model above, in terms of just a single equivalent-transport ellipsoid and its combined $m_{\textrm{equiv combined}}$ mass. That combined $m_{\textrm{equiv combined}}$ mass is measurable, but does not provide any explicit information about band warping, because all angular information is averaged out through the construction of the equivalent-transport ellipsoid. On the other hand, one may search for, and exploit band warping in order to produce shapes and relative configurations of equivalent ellipsoids that optimize transport properties.
 
Beyond transport theory and measurements, \textit{the angular dependence of band warping should be fully considered in the development of a general theory of optical transitions at critical points} and in corresponding experimental observations, particularly with regard to magneto-optical measurements.

\section{Conclusion}\label{Conclusion}

In this article, we have studied ``warping'' of electronic band structures, defined as not admitting second-order differentiability at critical points in $\bd{k}$-space. We have developed a general theory of band warping based on the introduction of an angular effective mass for \textit{radial expansions} at the critical point, which avoids the unwarranted assumption of a multi-dimensional quadratic expansion at that point in $\bd{k}$-space. We have demonstrated the key features of our theory by analyzing first-principles calculations and $\bd{k} \cdot \bd{p}$ models of warped and non-warped band structures in silicon. We then used our theory in the derivation and calculation of electronic transport properties and corresponding tensors. Our theory thus greatly enhances the heuristic value of the free electron model and sets the basis for exploiting band warping as a powerful tool to tailor transport properties in a variety of applications, starting with, but not limited to, thermoelectricity. 

We discovered that for any warped band structure at a critical point there is in principle an equivalent-transport ellipsoidal band that yields identical results from the standpoint of any transport property. Moreover, if two or more bands have the same degenerate energy at an extremum point or at equivalent multi-valley extrema, their equivalent-transport ellipsoids can be combined into a single equivalent-transport ellipsoid. These results have considerable import for the proper definition and measurement of effective masses of carriers in any crystal direction.

Our main results thus include: (1) a mathematically precise definition of warping in electronic band structure theory; (2) a clear definition and understanding of the origin and interplay of band non-parabolicity and warping in non-degenerate and degenerate bands at critical points; (3) a general theoretical and computational procedure to quantify warping in calculated or measured band structures; (4) a discussion of the transport application of the theory to multi-band and multi-valley models in anisotropic materials; (5) an explanation of the difficulty to detect band warping exclusively from electronic transport measurements, as a result of the formal existence of equivalent-transport ellipsoids; and (6) the possibility of drastic effects of band warping on electronic conductivity and thermopower tensors under certain conditions.

\acknowledgments
This work was supported by the Vitreous State Laboratory of The Catholic University of America. MF is grateful to Samsung's GRO program for partial support. We are grateful to three anonymous reviewers for their valuable comments and for pointing out both past and most recent literature of great relevance for our contribution.

%

\appendix
\section{Taylor Expansions}\label{appendix:Taylor}
Firstly, we should recall that the necessary and sufficient conditions for Taylor polynomial expansions are vastly more restrictive for functions of more than one variable, and secondly that for functions of more than one variable those conditions are even more restrictive for second- or higher-order differentiability than those for 1st-order differentiability.  Namely, for functions of a single variable, the Taylor polynomial expansion of $f(x)$ to order $n$ with Peano-form remainder merely requires that $f(x)$ is $n$-times differentiable at the expansion point $x_0$. However, for a function $f(x,y)$ of two (or more) variables, 1st-order differentiability requires the existence of a (tangent hyper-plane) differential as its linear approximation, plus a higher-order remainder that must vanish with the much more demanding condition of the two-dimensional limiting process, requiring independence of the limit on any path with which the extremum is approached. Beyond one-dimension, the existence of partial derivatives at an isolated point does not even imply continuity, let alone differentiability. For example, even a trivial two-dimensional function $f(x,y)$ that has constant value 1 along both the $x$- and $y$- axes and constant value 0 everywhere else is not continuous at (0,0) with the two-dimensional definition of limit, hence, it is not differentiable at (0,0) \textit{a fortiori}, despite the fact that both its $x$- and $y$- partial derivatives exist and are zero at (0,0). However, if a much stronger and sufficient condition is satisfied, namely that both partial derivatives exist in a whole neighborhood of a point $(x_0,y_0)$ and they are moreover continuous at that point, then differentiability of $f(x,y)$ is guaranteed at least at $(x_0,y_0)$. On the other hand, the converse still does not hold necessarily: $f(x,y)$ can admit a differential with corresponding partial derivatives at $(x_0,y_0)$ and yet those partial derivatives may not be continuous at $(x_0,y_0)$, or they may not even exist beyond that point.

Even greater restrictions intervene for a function $f(x,y)$ of two (or more) variables to attain second-order differentiability. For that it further becomes both necessary and sufficient that the Hessian matrix of second-order partial derivatives be continuous at the expansion point $(x_0,y_0)$. With that as a sufficient condition, Clairaut-Schwarz's theorem also insures symmetry of mixed partial derivatives at $(x_0,y_0)$ with respect to their ordering of differentiations. Under those conditions, the second-order differential, expressed as a quadratic form, osculates the graph of $f(x,y)$ to the extent of providing its full curvature at $(x_0,y_0)$ two-dimensionally. Such geometrical representation requires preservation of the real eigenvalues and eigenvectors of the Hessian matrix or the principal axes of the corresponding quadratic form under orthogonal transformations or rotations. Under more general curvilinear coordinate transformation, the two fundamental forms of differential geometry still provide the principal curvatures and the intrinsic Gaussian curvature of the surface.

Either the definition of second-order differentiability of $f(x,y)$ as having a \textit{Hessian matrix continuous at} $(x_0,y_0)$, or its consequence that the \textit{Hessian matrix at} $(x_0,y_0)$ \textit{must transform consistently under rotations about} $(x_0,y_0)$, can be used to decide in practice whether a two-dimensional quadratic expansion of $f(x,y)$ at $(x_0,y_0)$ is permissible or not. For numerical verifications, the latter transformation criterion is typically less demanding than the original defining criterion, which would require cumbersome computations of limiting processes in two (or more) dimensions. Let us then describe a possible practical implementation of the transformation criterion.  Given $f(x,y)$, we may evaluate its Hessian matrix $H$ at $(x_0,y_0)$ and we may find that it exists and it is symmetric with real eigenvalues $h_1$ and $h_2$. If that is not so, we can conclude immediately that $f(x,y)$ is not second-order differentiable. Otherwise, we may perform a rotation of the original Cartesian coordinate axes, using an orthogonal matrix $A$ and generating corresponding rotated coordinates $(x',y')$, in which the original function is expressed as $f'(x',y')=f(x,y)$. We may then obtain the Hessian matrix $H'$ corresponding to $f'(x',y')$ by evaluating the second-order partial derivatives of $f'(x',y')$ with respect to the rotated coordinate variables $(x',y')$. That Hessian matrix $H'$ may also exist and be symmetric, but $f(x,y)$ has no possibility of being second-order differentiable unless $H$ and $H'$ properly transform into one another, namely as $H'= A^T H A$, thus maintaining the same real eigenvalues $h_1=h_{1}'$ and $h_2=h_{2}'$ and the same orthogonal eigenvectors geometrically, namely those having algebraic components concordant with the coordinate transformation. Even if all that happens between $H$ and $H'$ for a particular rotation, in principle we should repeat the same process of verification for \textit{all} possible rotations, before concluding that $f(x,y)$ is second-order differentiable. In practice, it may be sufficient to repeat this transformation procedure only a few times to demonstrate, if we have a bit of insight or luck, that $f(x,y)$ is \textit{not} second-order differentiable, even if it admits apparently valid Hessian matrices in some particular coordinate systems. On the other hand, asserting that $f(x,y)$ \textit{is} definitely second-order differentiable may require more extensive testing or rigorous proof. 

It is straightforward to construct examples of elementary functions that are continuous and differentiable to 1st-order, but not differentiable to second-order at some points. Take for instance $f(x,y)$ defined as $f(0,0)=0$ and as a rational function $f(x,y)=xy(x^2-y^2)/(x^2+y^2)$ elsewhere. That function is 1st-order differentiable everywhere, with a vanishing differential at the origin. One can easily find that its second-order mixed partial derivatives have opposite values (+1 and -1) at $(0,0)$. Since they are discontinuous at $(0,0)$ Clairaut-Schwarz's symmetry theorem is not violated. The second-order double partial derivatives along each axis are instead zero at the origin. Thus we may still formally construct a Hessian matrix that has zero diagonal values and opposite off-diagonal values at $(0,0)$, but that cannot geometrically transform consistently under rotations. The corresponding quadratic Taylor polynomial would vanish, leaving $f(x,y)$ as a remainder that does not vanish faster than quadratically at the origin. So, no sensible quadratic Taylor expansion is possible at the origin for such an elementary function, and countless more of that sort.

Another elementary example that is associated with a prototypical physical situation of great interest, namely, \Eq{eqn:toymodel}, is provided by $g(x,y) = \sqrt{x^4 + y^4}$. That function is 1st-order differentiable everywhere, with a minimum at the origin. Therein, $g(x,y)$ has an apparently acceptable Hessian matrix $H$ of second-order partial derivatives, diagonal with two real and equal eigenvalues $h_1=h_2=2$. However, if we perform a rotation of $45^{\circ}$ of the Cartesian coordinate axes, the Hessian matrix $H'$ for the rotated coordinates of $g'(x',y')$ is still diagonal with two real and equal eigenvalues, but they have changed to $h'_1=h'_2=\sqrt{2}$. Such failure to maintain eigenvalue invariance tells us immediately that $H'$ must differ from the proper rotation $A^T H A$ of $H$. Hence, neither $H$ nor $H'$ can be ``bona fide'' Hessian matrices, although both of them may have looked like they were so individually. Moreover, $H$ has eigenvectors along the x- and y- axes, while $H'$ has eigenvectors along the $x=y$ and $x=-y$ diagonal axes. But a quadratic form correctly approximating $g(x,y)=g'(x',y')$ can admit only two orthogonal principal axes, not four different ones at increments of $45^{\circ}$ angles. The invalidity of Hessian matrices in other coordinate systems becomes most apparent for rotations of other arbitrary angles, since those yield generally anti-symmetric Hessian matrices with complex-conjugate pairs of eigenvalues. That means that \textit{mixed partial derivatives do not commute in almost all Cartesian coordinate systems}. So, we can readily conclude from our transformation criterion of Hessian matrices that $g(x,y)$ cannot admit a valid two-dimensional quadratic expansion at the origin, although its real and diagonal Hessian matrix $H$ may have looked initially suitable for that expansion, yielding ($x^2 + y^2$), which is nonsense however.

\section{Cartesian-to-polar coordinate transformations and changes of variables in transport tensor integrations}\label{appendix:transformation}
Rather than directly performing $\bd{k}$-space integrations in \Eq{eqn:boltzmanL} to evaluate matrix elements of transport tensors, it is possible and more convenient to perform a change of variables corresponding to a coordinate transformation from $(k_x,k_y,k_z)$ to $(E,\theta,\phi)$ as
\begin{align}
k_x &= \sqrt{\frac{E - E_c}{f_c(\theta,\phi)}} \sin \theta \cos \phi,\\
k_y &= \sqrt{\frac{E - E_c}{f_c(\theta,\phi)}} \sin \theta \sin \phi,\\
k_z &= \sqrt{\frac{E - E_c}{f_c(\theta,\phi)}} \cos \theta.
\end{align}
This transformation refers explicitly to a conduction band, although a corresponding transformation can be readily obtained for a valence band as well.

The determinant of the Jacobian matrix of these coordinate transformations for either conduction or valence bands is $J_{\stackrel{c}{v}}=\sqrt{|E - E_{\stackrel{c}{v}}|} \sin \theta / (2 \, |f_{\stackrel{c}{v}}(\theta,\phi)|^{3/2})$. Integration in $\dd k_x \dd k_y \dd k_z$ over the Brillouin zone can thus be replaced by integration in $\dd E \dd \theta \dd \phi$. However, energy eigenvalues are allowed only above $E_c$ and below $E_v$, since the density of states vanishes in the energy gap. We may thus split the integration over all energies into two (or more) parts, with one integral for each energy band.

There may be more complicated forms of $f(\theta,\phi)$ surfaces, having $(\theta,\phi)$ regions with alternating positive and negative $f(\theta,\phi)$ values, as typical of saddle points. However, this does not lead to major complications for our formalism. Namely, we only have to be mindful that integrations over positive $E - E_{0}$ values must be associated with regions of positive $f(\theta,\phi)$, while integrations over negative $E-E_0$ values must be associated with regions of negative $f(\theta,\phi)$. 

Let us then just set
\begin{equation}
k_r (E) = \left\{ \begin{array}{ll} \sqrt{\frac{(E - E_c)}{f_c (\theta, \phi)}} &\textrm{if} \, \, E \ge E_c, \\
\sqrt{\frac{(E_v - E)}{f_v (\theta, \phi)}} &\textrm{if} \, \, E \le E_v.
\end{array} \right.
\end{equation}
Straightforward calculations using the equations of this Appendix lead to \Eq{eqn:boltzkijnew} for the matrix elements of transport tensors in a two-band model.

\section{Anisotropic tensor forms of relaxation times}\label{appendixU}
Considering the possibility of anisotropic scattering in many-valley semiconductors,\cite{Herring1956, Ito1964} Bies \textit{et al.}\ \cite{Bies2002} have proposed to represent relaxation time as an anisotropic tensor of the form
\begin{equation}\label{eqn:BiesForm}
\underline{\bm{\tau}}_{n} = \tau_{n} (E_{n}(\bd{k})) \underline{\bm{U}}_{n},
\end{equation}
where $\underline{\bm{U}}_{n}$ is a constant tensor, independent of $\bd{k}$, and the subscript $n$ denotes an energy band or refers to a dependence on that band alone.

We notice at once that this tensorial form of relaxation time contains certain ambiguities and limitations. First of all, \Eq{eqn:BiesForm} is supposed to hold in a neighborhood of a critical point $\bd{k_0}$ where the energy band has a corresponding extremum. Suppose that the band is non-degenerate at $\bd{k_0}$ and thus admits a quadratic expansion therein. Physical expectations then require $\underline{\bm{U}}_{n}$ to be a symmetric tensor with the same principal axes as the quadratic form for the energy expansion. That was indeed the case proposed originally by Herring and Vogt,\cite{Herring1956} who considered ellipsoidal constant-energy surfaces and attributed to each surface a set of three different relaxation times, one for each of its three principal directions. In that case, the form\cite{Bies2002} of \Eq{eqn:BiesForm} is slightly less general, because it assumes the same energy dependence along each principal direction or eigenvector of $\underline{\bm{U}}_{n}$, although that energy dependence is multiplied by a different constant or eigenvalue of $\underline {\bm{U}}_{n}$. In any event, that is not a major limitation, and it may be adequate for multi-dimensional quadratic expansions that either Herring and Vogt\cite{Herring1956} or Bies \textit{et al.}\ \cite{Bies2002} have considered exclusively.

On the other hand, suppose that the band is degenerate at $\bd{k}_0$ and admits no multi-dimensional quadratic expansion therein, as is instead the focus of our investigation. Then it is unlikely that \Eq{eqn:BiesForm} can fully capture band-warping effects on relaxation times. For example, since now we have no principal directions, there is no intrinsic coordinate system to which the constant $\underline{\bm{U}}_{n}$ tensor, or its constant matrix elements $U_{n \, i,j}$, can be related physically, whatever symmetric or non-symmetric form may be arbitrarily presumed for $U_{n \, i,j}$. 

Notwithstanding this potentially major problem of associating a physically meaningful coordinate system to an arbitrary (and potentially meaningless) constant matrix $U_{n \, i,j}$, we intend to proceed here with the relaxation time form proposed\cite{Bies2002} in \Eq{eqn:BiesForm} and investigate which additional conditions or restrictions our theory of angular effective mass, encompassing band warping, may yet impose on that form. 

Let us first consider the energy dependence in the scalar factor $\tau_{n} (E_{n}(\bd{k}))$ of \Eq{eqn:BiesForm}. We may deal with that without problem through the energy integration in $\dd E$ at the level of \Eq{eqn:boltzkijnew}. However, any explicit energy dependence in the scalar factor $\tau_{n} (E_{n}(\bd{k}))$ should correspondingly modify the development of the $K_\alpha$ functions, which in the simplest version of our main text presume external factorization of energy-independent relaxation times in order to attain the ``universal'' form shown in \Eq{eqn:universalKn} and further discussed in Appendix\ \ref{appendixFD}.

Although allowing that energy dependence does not produce major complications for our formalism, there is really no major loss of generality in reducing \Eq{eqn:BiesForm} to the simpler form 
\begin{equation}\label{eqn:BiesForm1App}
\underline{\bm{\tau}}_{n} = \tau_{n} \, \underline{\bm{U}}_{n},
\end{equation}
where $\tau_{n}$ is assumed from now on to be a scalar constant for each band $n$.

The averaged-out tensors in \Eq{eqn:ctensors} then become
\begin{align}\label{eqn:ctensorsBies}
\underline{\bm{C}}_{n} &= C_{n \, i,j} = \nonumber\\&\tau_{n} \sum_{q} \int_{0}^{2 \pi} \! \! \! \int_{0}^{\pi} \frac{\hat{v}_{n,i}(\theta,\phi) U_{n \, j,q} \hat{v}_{n,q}(\theta,\phi)}{2 |f_{n}(\theta,\phi)|^{5/2}} \sin \theta \, \dd \theta \dd \phi,
\end{align}
which can be further expressed as
\begin{align}
\left[ \underline{\bm{C}}_n \right]_{i,j} = C_{n \, i,j} &= \sum_{q} \tau_{n} U_{n \, j,q} \mathcal{C}_{n \, i,q} \nonumber\\&= \tau_{n} \sum_{q} U_{n \, j,q} \mathcal{C}_{n \, q, i} = \tau_{n} \left[ (\underline{\bm{U}}_{n} \cdot \underline{\bm{\mathcal{C}}}_{n}) \right]_{j,i},
\end{align}
since
\begin{align}\label{eqn:scriptCAPP}
\left[ \underline{\bm{\mathcal{C}}}_{n} \right]_{i,j} &=  \mathcal{C}_{n \, i,j} = \nonumber\\&\int_{0}^{2 \pi} \! \! \! \int_{0}^{\pi} \frac{\hat{v}_{n i}(\theta,\phi) \hat{v}_{n j}(\theta,\phi)}{2 |f_{n}(\theta,\phi)|^{5/2}} \sin \theta \, \dd \theta \dd \phi
\end{align}
are manifestly symmetric matrices, coinciding with \Eq{eqn:scriptC} in our main text.

Since the product of two symmetric matrices is not necessarily symmetric, $\underline{\bm{C}}_n$ may not always turn out to be symmetric. However, assuming Onsager relations, we can prove that $\underline{\bm{C}}_n$ must always be symmetric as follows.

Based on the discussion after \Eq{eqn:onsagerrelations}, recall that the transport tensors $\underline{\pmb{\mathscr{L}}}^{(\alpha)}$ are always symmetric. Treating each band independently, we may regard $\underline{\pmb{\mathscr{L}}}^{(\alpha)}$ as resulting from sums over one-band models. In a one-band case, $\underline{\bm{C}}_n$ is proportional to the corresponding $\underline{\pmb{\mathscr{L}}}^{(\alpha)}_n$. Hence, we must have that $\underline{\bm{C}}_n$ are symmetric for all bands. QED.

Equivalently, we must have that
\begin{equation}
\left[ \underline{\bm{C}}_{n} \right]_{i,j} =  \left[ \tau_{n} \, \underline{\bm{U}}_{n} \cdot \underline{\bm{\mathcal{C}}}_{n} \right]_{j,i} = \left[ \tau_{n} \, \underline{\bm{U}}_{n} \cdot \underline{\bm{\mathcal{C}}}_{n} \right]_{i,j}.
\end{equation}
So, the product of the relaxation-time tensor, $\tau_{n} \underline{\bm{U}}_{n}$, and the $\underline{\bm{\mathcal{C}}}_{n}$ tensor must be symmetric.

Now, $\underline{\bm{\mathcal{C}}}_{n}$ is represented by a manifestly symmetric matrix in \Eq{eqn:scriptCAPP}. If $\underline{\bm{U}}_{n}$ is also represented by a symmetric matrix, it must be simultaneously diagonalizable with $\underline{\bm{\mathcal{C}}}_{n}$, because their product is symmetric, hence they must commute. On the other hand, if $\underline{\bm{U}}_{n}$ is \textit{not} symmetric, as may happen in case of band warping, the only condition that we can ultimately maintain is that the product $\underline{\bm{U}}_{n} \cdot \underline{\bm{\mathcal{C}}}_{n}$ must still be symmetric.

In our main text, we decided to side-step this more elaborate discussion by drastically assuming from \Eq{eqn:twobandL} onward that all $\underline{\bm{U}}_{n}$ tensors equal the identity, which amounts to the usual isotropic relaxation-time approximation for each band $n$. More technically, we could have retained the form of $\bm{\tau}_{n}$ given in \Eq{eqn:BiesForm1App} and all our major conclusions would have carried over essentially unchanged, except for formal but relatively straightforward generalizations, such as replacing \Eq{eqn:Ks} by
\begin{align}\label{eqn:Ksapp}
&\left[ \underline{\pmb{\mathscr{L}}}^{(\alpha)} \right]_{i,j} = \nonumber\\&\sum_{n = 1}^{N_b} (\textrm{m}_n)^{\alpha} \tau_{n}  K_{\alpha} (\textrm{m}_n \beta (E_{n} - \mu), \beta) \, \sum_q \mathcal{C}_{n \, i,q} U_{n \, j,q},
\end{align}
where the last factor, $\sum_q \mathcal{C}_{n \, i,q} U_{n \, j,q}$, still represents (a sum of) symmetric matrix elements in $(i,j)$.

\section{Asymptotic expansions of $K_{\alpha}$ and related transport functions}\label{appendixFD}
The ``universal'' functions $K_{\alpha}$ are defined in \Eq{eqn:universalKn} as
\begin{equation}
K_\alpha (s , \beta) = \frac{e^2 \sqrt{\me}}{2^{3/2} \pi^3 \hbar^3 \beta^{\alpha + 3/2}} \int_{s}^{\infty} \frac{x^\alpha (x - s)^{3/2} e^x}{(1+e^x)^2} \, \dd x.
\end{equation}

Fermi-Dirac (FD) integrals are defined as 
\begin{equation}
F_j(x) = \frac{1}{\Gamma(j+1)} \int_{0}^{\infty} \frac{t^j}{e^{t-x}+1} \, \dd t.
\end{equation}

Letting $A = e^2 \sqrt{\me} / (2^{3/2} \pi^3 \hbar^3)$, we can express with relatively straightforward calculations $K_{\alpha}$ in terms of FD integrals as
\begin{subequations}
\begin{align}
K_0 (s,\beta) &= \frac{A}{\beta^{3/2}} \frac{3}{2} \Gamma (3/2) F_{1/2} (-s), \\ 
K_1 (s,\beta) &= \frac{A}{\beta^{5/2}} \left( \frac{5}{2} \Gamma (5/2) F_{3/2} (-s) \right. \nonumber\\&\phantom{= \frac{A}{\beta^{5/2}} (} \left. + \frac{3}{2} s \Gamma (3/2) F_{1/2} (-s) \right), \\
K_2 (s,\beta) &= \frac{A}{\beta^{7/2}} \left( \frac{7}{2} \Gamma (7/2) F_{5/2} (-s) \right. \nonumber\\& \phantom{= \frac{A}{\beta^{7/2}} (} + 5 s \Gamma (5/2) F_{3/2} (-s) \nonumber\\  &\phantom{= \frac{A}{\beta^{7/2}} (}  \left. + s^2 \frac{3}{2} \Gamma (3/2) F_{1/2} (-s) \right). 
\end{align}
\end{subequations}

Asymptotic expansions of $K_{\alpha}$ for $s \gg 1$ (insulator limit) and for $s \ll -1$ (metallic limit) are given by
\begin{subequations}
\label{Appendixeqn:asymptoticapprox}
\begin{align}
K_0 (s,\beta) &\sim \frac{A}{\beta^{3/2}} \frac{3 \sqrt{\pi}}{4} e^{-s}, \nonumber\\& \quad s \gg 1 \, (\textrm{insulator}), \\ 
K_1 (s,\beta) &\sim \frac{A}{\beta^{5/2}} \frac{3 \sqrt{\pi}}{8} (5 + 2 s) e^{-s}, \nonumber\\& \quad s \gg 1 \, (\textrm{insulator}),\\
K_2 (s,\beta) &\sim \frac{A}{\beta^{7/2}} \frac{3 \sqrt{\pi}}{16} (35 + 4 s (5 + s)) e^{-s}, \nonumber\\& \quad s \gg 1 \, (\textrm{insulator}),\\
K_0 (s,\beta) &\sim \frac{A}{\beta^{3/2}} \left( (-s)^{3/2} + \frac{\pi^2}{8 \sqrt{-s}} \right), \nonumber\\& \quad s \ll -1 \, (\textrm{metal}), \\ 
K_1 (s,\beta) &\sim \frac{A}{\beta^{5/2}} \left( \frac{\pi^2}{2} \sqrt{-s} - \frac{7 \pi^4}{240 (-s)^{3/2}} \right), \nonumber\\& \quad s \ll -1 \, (\textrm{metal})\\
K_2 (s,\beta) &\sim \frac{A}{\beta^{7/2}} \left( \frac{\pi^2}{3} (-s)^{3/2} + \frac{7 \pi^2}{40 \sqrt{-s}} \right), \nonumber\\& \quad s \ll -1 \, (\textrm{metal}).
\end{align}
\end{subequations}

Using the expression for the ellipsoidal $\underline{\bm{\mathcal{C}}}$ given in \Eq{eqn:ellipsoidalC}, the asymptotic expansions for the eigenvalues of the conductivity and the thermopower tensors when the chemical potential $\mu$ lies sufficiently away from either band edge, on the order of $k_{\textrm{B}} T$, are given in the \textit{insulator limit} by
\begin{subequations}
\label{Appendixeqn:transporteqnsinsulating}
\begin{align}
\left[ \underline{\bm{\sigma}} \right]_{i,j} &= \left[ \underline{\pmb{\mathscr{L}}}^{(0)} \right]_{i,j} \nonumber\\ &\sim \frac{\tau e^2 \sqrt{m_1 m_2 m_3}}{\pi^{3/2} \sqrt{2} \hbar^3 m_{i} \beta^{3/2}} e^{-\textrm{m}_n \beta (E_n - \mu)} \, \delta_{i,j}, \\
\left[ \underline{\bm{S}} \right]_{i,j} &= \frac{-1}{T e} \left( \underline{\pmb{\mathscr{L}}}^{(0)} \right)^{-1} \cdot \underline{\pmb{\mathscr{L}}}^{(1)} \nonumber\\ &\sim -\textrm{m}_n \frac{k_\textrm{B}}{2 e} (5 + \textrm{m}_n 2 \beta (E_n - \mu)) \, \delta_{i,j}.
\end{align}
\end{subequations}

The corresponding asymptotic expansions for the eigenvalues of $\underline{\bm{\sigma}}$ and $\underline{\bm{S}}$ when the chemical potential $\mu$ lies within a given band are given in the \textit{metallic limit} by 
\begin{subequations}
\label{Appendixeqn:transporteqnsmetal}
\begin{align}
\left[ \underline{\bm{\sigma}} \right]_{i,j} &= \left[ \underline{\pmb{\mathscr{L}}}^{(0)} \right]_{i,j} \nonumber\\ &\sim \frac{\tau e^2 2^{3/2} \sqrt{m_1 m_2 m_3}}{3 \pi^{2} \hbar^3 m_{i}} (\textrm{m}_n (\mu - E_n))^{3/2} \, \delta_{i,j}, \\
\left[ \underline{\bm{S}} \right]_{i,j} &= \frac{-1}{T e} \left( \underline{\pmb{\mathscr{L}}}^{(0)} \right)^{-1} \cdot \underline{\pmb{\mathscr{L}}}^{(1)} \nonumber\\ &\sim \frac{k_\textrm{B} \pi^2}{2 e} \frac{1}{\beta (E_n - \mu)} \, \delta_{i,j}.
\end{align}
\end{subequations}

Here, $E_n$ represents either a band-edge minimum or maximum, and $\textrm{m}_n$ is either $+1$ for a conduction-like band, or $-1$ for a valence-like band.

It is relatively straightforward to derive the basic grand-canonical equation of state and invert it to express the chemical potential, $\mu$, in terms of the carrier density, $n$, in both the insulator and metallic limits.  For example, for a single conduction-like band,
\begin{subequations}
\label{Appendixeqn:mu-n}
\begin{align}
\beta ( E_0 - \mu ) &\sim - \log \left( \frac{\sqrt{2} \hbar^3 \pi^{3/2} \beta^{3/2}}{\sqrt{m_x m_y m_z}} n \right), \nonumber\\& \quad \beta (E_0 - \mu) \gg 1 \, (\textrm{insulator}),\\
\beta ( E_0 - \mu ) &\sim - \frac{3^{2/3} \hbar^2 \pi^{4/3} \beta}{2 m_x^{1/3} m_y^{1/3} m_z^{1/3}} n^{2/3}, \nonumber\\& \quad \beta (E_0 - \mu) \ll -1 \, (\textrm{metal}).
\end{align}
\end{subequations}
One can thus obtain the corresponding asymptotic expressions for the eigenvalues of $\underline{\bm{\sigma}}$ and $\underline{\bm{S}}$ in terms of $n$, namely 
\begin{subequations}
\label{Appendixeqn:cutlermott}
\begin{align}
\left[ \underline{\bm{\sigma}} \right]_{i,j} &\sim \frac{e^2 \tau n}{m_i} \delta_{i,j}, \nonumber\\& \quad \beta (E_0 - \mu) \gg 1 \, (\textrm{insulator}), \\ 
\left[ \underline{\bm{S}} \right]_{i,j} &\sim \frac{-k_{\textrm{B}}}{2 e} \left( 5 - 2 \log \left( \frac{\sqrt{2} \hbar^3 \pi^{3/2} \beta^{3/2}}{\sqrt{m_x m_y m_z}} n \right) \right) \delta_{i,j}, \nonumber\\& \quad \beta (E_0 - \mu) \gg 1 \, (\textrm{insulator}),\\
\left[ \underline{\bm{\sigma}} \right]_{i,j} &\sim \frac{e^2 \tau n}{m_i} \delta_{i,j}, \nonumber\\& \quad \beta (E_0 - \mu) \ll -1 \, (\textrm{metal}), \\ 
\left[ \underline{\bm{S}} \right]_{i,j} &\sim - \frac{k_{\textrm{B}} \pi^{2/3} m_x^{1/3} m_y^{1/3} m_z^{1/3}}{e 3^{2/3} \hbar^2 \beta n^{2/3}} \delta_{i,j}, \nonumber\\& \quad \beta (E_0 - \mu) \ll -1 \, (\textrm{metal}).
\end{align}
\end{subequations}
In particular, in the metallic limit, the relation between the Seebeck coefficient and the electrical conductivity reduces to the expression derived from the Cutler-Mott formula for isotropic bands,\cite{Cutler, ParkerPRL2013, Chen2013, Filippetti2012, Snyder2008complex} namely
\begin{equation}
S=-\frac{k_{\textrm{B}}}{e \hbar^2} m^{*} \left( \frac{\pi}{3 n} \right)^{2/3}.
\end{equation}
Our formalism further allows extensions of the Cutler-Mott formula to warped, anisotropic, and multi-band cases in both the insulator and metallic limits. 
\end{document}